\documentclass{jfm}
 
\usepackage{natbib}  	
\usepackage{amsmath}
\usepackage{amssymb}
\usepackage{graphicx,color,natbib,enumitem}
\usepackage{amsfonts,graphicx,wrapfig}
\usepackage{xcolor}
\usepackage{hyperref}
\definecolor{Gray}{gray}{0.9}
\definecolor{LightCyan}{rgb}{0.88,1,1}
\usepackage{verbatim}
\usepackage{colortbl}
\usepackage{tikz}

\newcommand{\q}[1]{\textcolor{red}{#1}}

\newcolumntype{L}{>{\arraybackslash}m{1.5 cm}}

\shorttitle{Rough-wall drag prediction using machine learning}

\shortauthor{M. Aghaei Jouybari et al.}

\author{Mostafa Aghaei Jouybari\aff{1}\corresp{\email{aghaeijo@msu.edu}}, Junlin Yuan\aff{1}, Giles J. Brereton\aff{1} \and Michael S. Murillo\aff{2}  }

\affiliation{\aff{1}Department of Mechanical Engineering, Michigan State University, East Lansing, MI 48824, USA
\aff{2}Department of Computational Mathematics, Science and Engineering, Michigan State University, East Lansing, MI 48824, USA
}

\title{Data-driven prediction of the equivalent sand-grain height in rough-wall turbulent flows}

\begin{document}
\maketitle

\begin{abstract}
This paper investigates a long-standing question about the effect of surface roughness on turbulent flow: what is the equivalent roughness sand-grain height for a given roughness topography? Deep Neural Network (DNN) and Gaussian Process Regression (GPR) machine learning approaches are used to develop a high-fidelity prediction approach of the Nikuradse equivalent sand-grain height $k_s$   for turbulent flows over a wide variety of different rough surfaces. To this end, 45 surface geometries were generated and the flow over them simulated at $\hbox{Re}_\tau=1000$ using direct numerical simulations. These surface geometries differed significantly in moments of surface height fluctuations, effective slope, average inclination, porosity and degree of randomness. Thirty of these surfaces were considered fully-rough and they were supplemented with experimental data for fully-rough flows over 15 more surfaces available from previous studies. The DNN and GPR methods predicted $k_s$ with an average error of less than 10\%  and a maximum error of less than 30\%, which appears to be significantly more accurate than existing prediction formulas. They also identified the surface porosity and the effective slope of roughness in the spanwise direction as important factors in drag prediction.
\end{abstract}

\section{Introduction}\label{sec:intro}

At sufficiently high Reynolds numbers all surfaces are hydrodynamically rough, as is almost always the case for flows past the surfaces of naval vehicles. Reviews of roughness effects on wall-bounded turbulent flows are provided by \citet{RaupachAR91}  and \citet{Jimenez04}. The most important effect of surface roughness in engineering applications is an increase in the hydrodynamic drag~\citep{FlackF18a}, which is due predominantly to the  pressure drag generated by the small-scale recirculation regions associated with individual roughness protuberances.

For the foreseeable future, the most practical approach to making predictive flow calculations for many realistic applications is to use engineering one-point closures of turbulence, such as two-equation turbulent eddy-viscosity models to the Reynolds-averaged Navier-Stokes (RANS) equations. Existing rough-wall corrections to this type of closure typically model the increase in hydrodynamic drag on a single length scale---the equivalent sand-grain height \citep{Nikuradse33e} $k_s$---without physically resolving the surface or changing the governing equations. In the fully rough flow regime, where the wall friction depends on the roughness alone and is independent of the Reynolds number, $k_s$ was observed to quantify the increase in hydrodynamic drag through the empirical relation with the roughness function, $\Delta U^+$ (defined as the offset of the log-linear velocity profile of a rough-wall flow relative to that of a smooth-wall one):
\begin{equation}
\label{eq:ks_def}
    \Delta U^+=\frac{1}{\kappa}\ln{k_s^+}-3.5,
\end{equation}
where $\kappa=0.41$ is von K\'arm\'an's constant and $+$ represents  normalization in wall units.

A universal length scale (e.g. $k_s$ in Nikuradse's relation, or $\epsilon$ in the Moody diagram \citep{Moody44}) that can predict accurately the surface drag coefficient is not known a priori and does not appear to be equivalent to any single geometrical length scale, such as an average or a root-mean-square (rms) of roughness height  \citep{FlackF18a}. It is also well-established that $k_s$ can depend on many geometrical parameters such as the effective slope \citep{NapoliNAD08a,YuanP14a} and the skewness of the roughness height distribution \citep{FlackS10}. 
Readers are referred to \citet{FlackS10} and  \citet{Bons02}   for extensive reviews on this topic.  Empirical expressions for $k_s$ based on a small number of geometrical roughness parameters include, among others:
\begin{equation}
\label{eq:empmdls}
    k_s=c_1k_{avg}\big(\alpha_{rms}^2+c_2\alpha_{rms}\big)\ ,\quad
    k_s=c_1k_{avg}\Lambda_s^{c_2}\ ,\quad \hbox{and}\quad
    k_s=c_1k_{rms}\big(1+S_k\big)^{c_2},
\end{equation}
proposed by \citet{BonsEtAl01}, \citet{vanRijEtAl02}
and \citet{FlackS10} respectively. Here $k_{avg}$ is the average height, $\alpha$ is the local streamwise slope angle and $\Lambda_s=\big(S/S_f\big)\big(S_f/S_s\big)^{-1.6}$   (where $S$, $S_f$, $S_s$  are, respectively, the platform area, the total frontal area, and the total windward wetted area  of the roughness) while  $k_{rms}$ and  $S_k$ are the rms and skewness of the roughness height fluctuations, and $c_1$ and $c_2$ are constants.

The hydrodynamic lengthscale $k_s$ appears to be correlated with different sets of geometrical parameters for each type of rough surface and no universal correlation currently exists for flow over surfaces of arbitrary roughness. For example, for synthetic roughness comprising closely packed pyramids \citep{SchultzF09} and random sinusoidal waves \citep{NapoliNAD08a}, it has been shown that $k_s$ scales on the effective slope when  the surface slope is gentle (i.e. within the `waviness' regime), whereas the skewness and rms height, but \textit{not} slope magnitude, become important when the slope is steeper (i.e. within the `roughness' regime). The boundary between these two regimes has been shown to be surface dependent \citep{YuanP14a}.

Some more recent studies of $k_s$ correlations are summarized below. \cite{ThakkarTBS17} carried out DNS of transitionally-rough  turbulent flows for different irregular roughness topographies. They found that the roughness function is influenced by solidity, skewness, the streamwise correlation length scale and the rms of roughness height. \cite{FlackFSB19a} performed several experiments to systematically investigate the effects of the  skewness and amplitude of roughness height on the skin friction. They found that the rms and skewness of roughness height fluctuations are important scaling parameters for prediction of roughness function; however,  the surfaces with positive, negative and zero skewness values needed different correlations. Also, \cite{ChanCMCHO15} simulated turbulent pipe flows over sinusoidal roughness geometries and confirmed strong dependence of roughness function on the   average height and streamwise effective slope. 

In previous studies, the small number of roughness parameters used to devise $k_s$ correlations tended to limit their application to a narrow range of surface roughnesses.
Since it appears that many geometrical parameters, such as porosity, moments of roughness height (e.g. rms, skewness and kurtusis), effective slope, and surface inclination angle might affect $k_s$, it is useful to employ a data science approach suited to modeling large multi-variate/multi-output systems.

Specifically, we use Machine Learning (ML) to explore $k_s$-prediction approaches that depend on a large set of surface-topographical parameters, with the expectation that the resulting models may be applied accurately to a wider range of surfaces. Since the prediction of $k_s$ from surface topography is essentially a labeled regression problem, supervised ML operations were performed using Deep Neural Networks (DNN) and Gaussian Process Regressions (GPR). Both methods are explained thoroughly in section \ref{sec:results}.  Readers are referred to the monograph by \citet{Rasmussen06} and the review provided by \citet{LeCun15} for detailed descriptions of these methods. 

An initial ensemble of 60 sets of data on $k_s$ as a function of topographical parameters---45 direct numerical simulation (DNS) results and 15 experimental results---was considered. All experimental data sets are fully rough, and of the DNS data, 30 are considered fully-rough flows; all fully-rough cases were used for ML training and testing. To the best of our knowledge, this ensemble of roughness geometries is the most extensive used  for developing a  $k_s$-prediction method.

In this paper, we first present the governing equations, solution methodologies, simulation parameters and different roughness topographies and then discuss the  post-processed DNS results used to calculate $k_s$ for each surface. Finally, we describe the ML models, their predictions of $k_s$ and their uncertainty.

\section{Problem formulation}\label{sec:method_ks}  
\subsection{Governing equations}
The governing equations of incompressible continuity and linear momentum---the Navier-Stokes (NS) equations---for a constant-property Newtonian fluid, were solved by DNS. These equations are written in indicial notation as
 \begin{subequations}
\begin{align}
    \frac{\partial u_i}{\partial x_i} &=0, \label{eq:gov1} \\
  \frac{\partial u_i}{\partial t}+ \frac{\partial u_i u_j}{\partial x_j} &=-\frac{\partial P}{\partial x_i}
   +\nu\frac{\partial^2 u_i}{\partial x_j \partial x_j} + F_i,\>\>  
   \label{eq:gov2}
\end{align}
\end{subequations}
where $i,j=1,2,3$, $x_1, x_2$ and $x_3$ (or $x, y, z$) are the streamwise, wall-normal and spanwise coordinates, with corresponding velocity components of $u_1, u_2$ and $u_3$ (or $u, v, w$) and $P$ is defined as $p/\rho$, where $p$ is the pressure and $\rho$ is the fluid density; $\nu$ is the kinematic viscosity. An immersed boundary method \citep{YuanP14} was used to enforce  the fine-grained roughness boundary conditions on a non-conformal Cartesian grid. The corresponding body force $F_i$ is added to the the right hand side of the momentum equations to impose a no-slip boundary condition at the fluid-roughness interface. To solve the equations, second-order central differencing was used for spatial discretizations and second-order Adams-Bashforth semi-implicit time advancement was employed. The numerical solver was parallelized using a message passing interface (MPI) method \citep{Keating04}.

A double-averaging decomposition \citep{RaupachS82} was used to resolve turbulent and dispersive  components of flow variables in the presence of roughness. In this decomposition,
any instantaneous flow variable $\theta$ may be decomposed into three components, as
\begin{equation}
  \theta(\boldsymbol{x},t)=\big<\overline{\theta}\big>(y)+\theta'(\boldsymbol{x},t)+\widetilde{\theta}(\boldsymbol{x})
  \label{eq:decom}
\end{equation} 
where the time-averaging operator is $\overline{\theta}$ and the intrinsic spatial-averaging operator is $\big<\theta\big>=1/A_f\int_{x,z}\theta dA$ (and $A_f(y)$ is the area occupied by fluid at an elevation $y$). The Reynolds and dispersive fluctuating components are then $\theta'=\theta-\overline{\theta}$ and $\widetilde{\theta}=\overline{\theta}-\big<\overline{\theta}\big>$ respectively. $\langle \overline{\theta}\rangle$ is called the double-averaged component.

The wall shear stress (including both viscous and pressure drag contributions on a rough wall) was determined by integrating the time-averaged immersed boundary method  body force in the $x$-direction $F_1$ as
\begin{equation}
\tau_w = \frac{\rho}{L_x L_z}\int_\mathcal{V} \overline{F_1}(x,y,z) dxdydz,
\end{equation}
where $\mathcal{V}$ represents the simulation domain volume below the roughness crest and $L_{x_i}$ is the domain length in the $x_i$ direction. Readers are referred to  \citet{YuanP14,YuanP14b} for details of the implementation and validation of the immersed boundary method and the $\tau_w$  calculation.

\subsection{Surface roughness}

\begin{figure}
   \centerline{\includegraphics[width=1\textwidth,trim={0 0cm 0 0cm},clip]{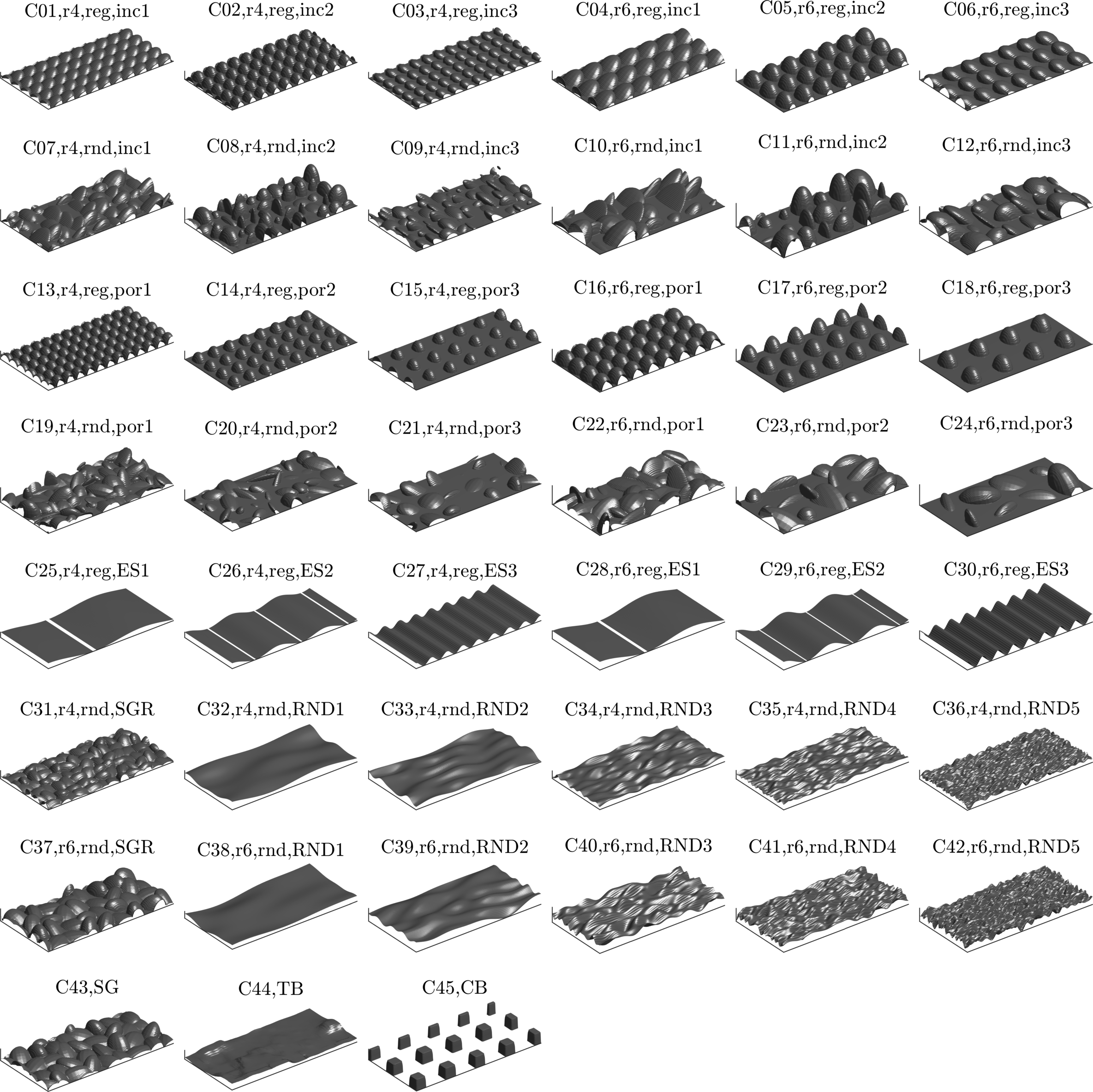}}
   \caption{Roughness geometries---each plot is a section of size $\delta \times 0.5 \delta$ in the $x$-$z$ plane. Cases C43 to C45 are from    simulations  with regular domain sizes  \citep{YuanP14a,Aghaei-JouybariABY19a}.
   }
\label{fig:vof}
\end{figure}

 \begin{table}
   \begin{center}{\tiny 
     \def~{\hphantom{10}} 
     \begin{tabular}{l c c c c c c c c c c c c c }
 {\footnotesize Case name}	 &   {\footnotesize $k_{avg}$}	 & {\footnotesize	$k_c$} & {\footnotesize	$k_t$}	 & {\footnotesize	$k_{rms}$}	 & {\footnotesize	$R_a$}	 & 	{\footnotesize $I_x$}  & {\footnotesize $I_z$}	 & 	{\footnotesize $P_o$}	 & {\footnotesize	$E_x$}	 & 	{\footnotesize $E_z$}	 & 	{\footnotesize $S_k$}	 & 	{\footnotesize $K_u$}    &  {\footnotesize $k_s$ }\\ 
\rowcolor{LightCyan} {\footnotesize C01,r4,reg,inc1}	 & 	0.026	 & 	0.043	 & 	0.043	 & 	0.013	 & 	0.011	 & 	-0.801	 & 	-0.089	 & 	0.535	 & 	0.584	 & 	0.510	 & 	-0.544	 & 	2.177	 & 		 \\ 
{\footnotesize C02,r4,reg,inc2}	                         & 	0.030	 & 	0.059	 & 	0.059	 & 	0.021	 & 	0.019	 & 	0.012	 & 	0.032	 & 	0.609	 & 	1.029	 & 	0.562	 & 	-0.265	 & 	1.597	 & 		 \\ 
\rowcolor{LightCyan} {\footnotesize C03,r4,reg,inc3}	 & 	0.025	 & 	0.043	 & 	0.043	 & 	0.013	 & 	0.011	 & 	0.821	 & 	-0.078	 & 	0.537	 & 	0.600	 & 	0.485	 & 	-0.459	 & 	2.052	 & 			 \\ 
{\footnotesize C04,r6,reg,inc1}	                         & 	0.032	 & 	0.064	 & 	0.064	 & 	0.022	 & 	0.019	 & 	-0.978	 & 	0.016	 & 	0.597	 & 	0.595	 & 	0.590	 & 	-0.167	 & 	1.601	 & 	 0.064		 \\ 
\rowcolor{LightCyan} {\footnotesize C05,r6,reg,inc2}	 & 	0.038	 & 	0.088	 & 	0.088	 & 	0.033	 & 	0.030	 & 	0.025	 & 	0.064	 & 	0.654	 & 	0.916	 & 	0.643	 & 	0.109	 & 	1.436	 & 	0.124		 \\ 
{\footnotesize C06,r6,reg,inc3}	                         & 	0.031	 & 	0.064	 & 	0.064	 & 	0.022	 & 	0.019	 & 	0.955	 & 	0.121	 & 	0.599	 & 	0.588	 & 	0.558	 & 	-0.087   & 	1.590	 & 	0.059		 \\ 
\rowcolor{LightCyan} {\footnotesize C07,r4,rnd,inc1}	 & 	0.025	 & 	0.086	 & 	0.084	 & 	0.022	 & 	0.019	 & 	-0.860	 & 	0.033	 & 	0.774	 & 	0.511	 & 	0.559	 & 	0.560	 & 	2.244	 & 	0.136		 \\ 
{\footnotesize C08,r4,rnd,inc2}	                         & 	0.027	 & 	0.116	 & 	0.115	 & 	0.030	 & 	0.025	 & 	-0.007	 & 	0.048	 & 	0.819	 & 	0.861	 & 	0.604	 & 	0.870	 & 	2.627	 & 	0.322		 \\ 
\rowcolor{LightCyan} {\footnotesize C09,r4,rnd,inc3}	 & 	0.025	 & 	0.083	 & 	0.081	 & 	0.021	 & 	0.018	 & 	0.829	 & 	0.002	 & 	0.753	 & 	0.517	 & 	0.482	 & 	0.514	 & 	2.292	 & 	0.131		 \\ 
{\footnotesize C10,r6,rnd,inc1}	                         & 	0.026	 & 	0.125	 & 	0.120	 & 	0.030	 & 	0.025	 & 	-0.957	 & 	-0.019	 & 	0.835	 & 	0.498	 & 	0.578	 & 	0.967	 & 	2.874	 & 	0.269		 \\ 
\rowcolor{LightCyan} {\footnotesize C11,r6,rnd,inc2}	 & 	0.033	 & 	0.172	 & 	0.169	 & 	0.044	 & 	0.037	 & 	0.076	 & 	0.138	 & 	0.842	 & 	0.758	 & 	0.543	 & 	1.150	 & 	3.176	 & 	0.536		 \\ 
{\footnotesize C12,r6,rnd,inc3}	                         & 	0.032	 & 	0.127	 & 	0.121	 & 	0.032	 & 	0.027	 & 	0.923	 & 	0.032	 & 	0.784	 & 	0.508	 & 	0.471	 & 	0.758	 & 	2.642	 & 	0.272		 \\ 
\rowcolor{LightCyan} {\footnotesize C13,r4,reg,por1}	 & 	0.038	 & 	0.059	 & 	0.059	 & 	0.018	 & 	0.015	 & 	0.024	 & 	0.067	 & 	0.498	 & 	1.043	 & 	0.523	 & 	-0.820	 & 	2.508	 & 			    \\ 
{\footnotesize C14,r4,reg,por2}	                         & 	0.018	 & 	0.059	 & 	0.059	 & 	0.022	 & 	0.020	 & 	0.021	 & 	0.038	 & 	0.776	 & 	0.613	 & 	0.456	 & 	0.708	 & 	1.840	 & 	0.141		 \\ 
\rowcolor{LightCyan} {\footnotesize C15,r4,reg,por3}	 & 	0.010	 & 	0.059	 & 	0.059	 & 	0.019	 & 	0.014	 & 	0.022	 & 	0.063	 & 	0.877	 & 	0.334	 & 	0.253	 & 	1.646	 & 	4.094	 & 	0.157		 \\ 
{\footnotesize C16,r6,reg,por1}	                         & 	0.051	 & 	0.089	 & 	0.089	 & 	0.030	 & 	0.026	 & 	0.041	 & 	0.149	 & 	0.529	 & 	1.137	 & 	0.534	 & 	-0.538	 & 	1.873	 & 	0.077		 \\ 
\rowcolor{LightCyan} {\footnotesize C17,r6,reg,por2}	 & 	0.022	 & 	0.089	 & 	0.089	 & 	0.031	 & 	0.027	 & 	0.041	 & 	0.080	 & 	0.801	 & 	0.537	 & 	0.403	 & 	0.982	 & 	2.308	 & 	0.260		 \\ 
{\footnotesize C18,r6,reg,por3}	                         & 	0.013	 & 	0.089	 & 	0.089	 & 	0.026	 & 	0.020	 & 	0.057	 & 	0.126	 & 	0.886	 & 	0.307	 & 	0.230	 & 	1.849	 & 	4.839	 & 	0.247		 \\ 
\rowcolor{LightCyan} {\footnotesize C19,r4,rnd,por1}	 & 	0.027	 & 	0.112	 & 	0.108	 & 	0.021	 & 	0.017	 & 	0.025	 & 	-0.107	 & 	0.806	 & 	0.487	 & 	0.486	 & 	0.732	 & 	3.422	 & 	0.158		 \\ 
{\footnotesize C20,r4,rnd,por2}	                         & 	0.013	 & 	0.095	 & 	0.087	 & 	0.017	 & 	0.014	 & 	0.032	 & 	-0.646	 & 	0.896	 & 	0.311	 & 	0.323	 & 	1.343	 & 	4.126	 & 	0.106		 \\ 
\rowcolor{LightCyan} {\footnotesize C21,r4,rnd,por3}	 & 	0.009	 & 	0.098	 & 	0.094	 & 	0.016	 & 	0.012	 & 	0.321	 & 	-0.741	 & 	0.929	 & 	0.219	 & 	0.233	 & 	2.168	 & 	7.728	 & 	0.103		 \\ 
{\footnotesize C22,r6,rnd,por1}	                         & 	0.035	 & 	0.139	 & 	0.139	 & 	0.029	 & 	0.024	 & 	-0.070	 & 	-0.245	 & 	0.791	 & 	0.456	 & 	0.499	 & 	0.591	 & 	2.830	 & 	0.277		 \\ 
\rowcolor{LightCyan} {\footnotesize C23,r6,rnd,por2}	 & 	0.017	 & 	0.123	 & 	0.111	 & 	0.025	 & 	0.020	 & 	-0.672	 & 	-0.841	 & 	0.885	 & 	0.305	 & 	0.325	 & 	1.467	 & 	4.347	 & 	0.175		 \\ 
{\footnotesize C24,r6,rnd,por3}	                         & 	0.014	 & 	0.152	 & 	0.145	 & 	0.027	 & 	0.019	 & 	0.189	 & 	-0.056	 & 	0.926	 & 	0.254	 & 	0.257	 & 	2.371	 & 	8.740	 & 	0.260		 \\ 
\rowcolor{LightCyan} {\footnotesize C25,r4,reg,ES1}	     & 	0.020	 & 	0.040	 & 	0.040	 & 	0.014	 & 	0.013	 & 	0.046	 & 	0.006	 & 	0.510	 & 	0.106	 & 	0.009	 & 	-0.032	 & 	1.503	 & 			     \\ 
{\footnotesize C26,r4,reg,ES2}	                         & 	0.021	 & 	0.040	 & 	0.040	 & 	0.014	 & 	0.013	 & 	0.039	 & 	-0.001	 & 	0.510	 & 	0.212	 & 	0.020	 & 	-0.071	 & 	1.505	 & 	0.065		 \\ 
\rowcolor{LightCyan} {\footnotesize C27,r4,reg,ES3}	     & 	0.023	 & 	0.040	 & 	0.040	 & 	0.014	 & 	0.012	 & 	0.006	 & 	-0.023	 & 	0.510	 & 	0.609	 & 	0.032	 & 	-0.214	 & 	1.544	 & 			   \\ 
{\footnotesize C28,r6,reg,ES1}	                         & 	0.030	 & 	0.059	 & 	0.059	 & 	0.021	 & 	0.019	 & 	0.044	 & 	0.018	 & 	0.504	 & 	0.158	 & 	0.015	 & 	-0.031	 & 	1.499	 & 	0.071		 \\ 
\rowcolor{LightCyan} {\footnotesize C29,r6,reg,ES2}	     & 	0.031	 & 	0.059	 & 	0.059	 & 	0.021	 & 	0.019	 & 	0.028	 & 	-0.069	 & 	0.504	 & 	0.316	 & 	0.022	 & 	-0.071	 & 	1.503	 & 	0.112		 \\ 
{\footnotesize C30,r6,reg,ES3}	                         & 	0.034	 & 	0.059	 & 	0.059	 & 	0.020	 & 	0.018	 & 	0.015	 & 	-0.069	 & 	0.505	 & 	0.917	 & 	0.048	 & 	-0.203	 & 	1.543	 & 	0.064		 \\ 
\rowcolor{LightCyan} {\footnotesize C31,r4,rnd,SGR}	     & 	0.025	 & 	0.059	 & 	0.059	 & 	0.011	 & 	0.009	 & 	0.104	 & 	-0.039	 & 	0.648	 & 	0.370	 & 	0.398	 & 	0.378	 & 	2.784	 & 	0.049		 \\ 
{\footnotesize C32,r4,rnd,RND1}	                         & 	0.040	 & 	0.075	 & 	0.072	 & 	0.013	 & 	0.010	 & 	0.117	 & 	0.108	 & 	0.479	 & 	0.068	 & 	0.169	 & 	-0.069	 & 	2.991	 & 			    \\ 
\rowcolor{LightCyan} {\footnotesize C33,r4,rnd,RND2}	 & 	0.041	 & 	0.088	 & 	0.084	 & 	0.013	 & 	0.011	 & 	0.109	 & 	0.078	 & 	0.553	 & 	0.117	 & 	0.308	 & 	0.004	 & 	2.763	 & 			   \\ 
{\footnotesize C34,r4,rnd,RND3}	                         & 	0.042	 & 	0.080	 & 	0.071	 & 	0.010	 & 	0.008	 & 	0.070	 & 	0.051	 & 	0.508	 & 	0.175	 & 	0.458	 & 	-0.002	 & 	3.031	 & 			    \\ 
\rowcolor{LightCyan} {\footnotesize C35,r4,rnd,RND4}	 & 	0.043	 & 	0.077	 & 	0.066	 & 	0.008	 & 	0.007	 & 	0.039	 & 	0.042	 & 	0.488	 & 	0.218	 & 	0.558	 & 	0.013	 & 	2.941	 & 			    \\ 
{\footnotesize C36,r4,rnd,RND5}	                         & 	0.045	 & 	0.084	 & 	0.067	 & 	0.009	 & 	0.007	 & 	0.035	 & 	0.037	 & 	0.535	 & 	0.378	 & 	0.841	 & 	0.075	 & 	3.018	 & 			    \\ 
\rowcolor{LightCyan} {\footnotesize C37,r6,rnd,SGR}	     & 	0.037	 & 	0.088	 & 	0.088	 & 	0.018	 & 	0.015	 & 	0.312	 & 	0.180	 & 	0.640	 & 	0.428	 & 	0.463	 & 	0.323	 & 	2.686	 & 	0.109		 \\ 
{\footnotesize C38,r6,rnd,RND1}	                         & 	0.060	 & 	0.106	 & 	0.091	 & 	0.016	 & 	0.012	 & 	0.045	 & 	0.028	 & 	0.444	 & 	0.077	 & 	0.183	 & 	-0.220	 & 	3.258	 & 			    \\ 
\rowcolor{LightCyan} {\footnotesize C39,r6,rnd,RND2}	 & 	0.061	 & 	0.098	 & 	0.095	 & 	0.012	 & 	0.009	 & 	0.111	 & 	0.057	 & 	0.400	 & 	0.108	 & 	0.285	 & 	-0.020	 & 	3.267	 & 			    \\ 
{\footnotesize C40,r6,rnd,RND3}                     	 & 	0.064	 & 	0.121	 & 	0.112	 & 	0.016	 & 	0.013	 & 	0.061	 & 	0.022	 & 	0.512	 & 	0.280	 & 	0.760	 & 	0.037	 & 	2.977	 & 	0.050		 \\ 
\rowcolor{LightCyan} {\footnotesize C41,r6,rnd,RND4}	 & 	0.065	 & 	0.130	 & 	0.130	 & 	0.015	 & 	0.012	 & 	0.045	 & 	0.037	 & 	0.546	 & 	0.374	 & 	0.989	 & 	0.028	 & 	3.036	 & 			    \\ 
{\footnotesize C42,r6,rnd,RND5}	                         & 	0.068	 & 	0.118	 & 	0.116	 & 	0.013	 & 	0.010	 & 	0.037	 & 	0.025	 & 	0.503	 & 	0.547	 & 	1.204	 & 	0.052	 & 	2.933	 & 			    \\ 
\rowcolor{LightCyan} {\footnotesize C43,SG}	             & 	0.036	 & 	0.089	 & 	0.087	 & 	0.017	 & 	0.014	 & 	0.288	 & 	0.156	 & 	0.649	 & 	0.425	 & 	0.441	 & 	0.476	 & 	2.970	 & 	0.093		 \\ 
{\footnotesize C44,TB}	                                 & 	0.055	 & 	0.125	 & 	0.088	 & 	0.018	 & 	0.014	 & 	0.007	 & 	-0.006	 & 	0.569	 & 	0.097	 & 	0.081	 & 	0.200	 & 	3.493	 & 	0.024		 \\ 
\rowcolor{LightCyan} {\footnotesize C45,CB}	             & 	0.010	 & 	0.070	 & 	0.070	 & 	0.023	 & 	0.016	 & 	0.420	 & 	0.508	 & 	0.878	 & 	0.249	 & 	0.247	 & 	2.101	 & 	5.569	 & 	0.150	 \\ 
{\footnotesize  C46,r4,rnd,por3,FS}	                     & 	0.009	 & 	0.098	 & 	0.094	 & 	0.016	 & 	0.012	 & 	0.321	 & 	-0.715	 & 	0.929	 & 	0.219	 & 	0.234	 & 	2.168	 & 	7.728	 & 	0.104		 \\ 
\rowcolor{LightCyan}  {\footnotesize E01,16,2}	 	     & 	0.138	 & 	0.261    & 0.254	 & 	0.020	 & 	0.016	 & 	-0.005	 & 	0.011	 & 	0.472	 & 	0.720	 & 	0.835	 & 	-0.711	 & 	3.843	 & 	0.052      \\
 {\footnotesize E02,16,3}	 	                         & 	0.143	 & 	0.252	 & 0.252    & 	0.021	 & 	0.016	 & 	-0.021	 & 	0.010	 & 	0.432	 & 	0.740	 & 	0.868	 & 	-0.338	 & 	3.159	 & 	0.050    \\
\rowcolor{LightCyan}  {\footnotesize E03,16,7}	    	 & 	0.133	 & 	0.365    & 0.254	 & 	0.019	 & 	0.014	 & 	-0.038   & 	0.000	 & 	0.638	 & 	0.618	 & 	0.705	 & 	-1.169	 & 	5.292	 & 	0.058     \\
 {\footnotesize E04,16,8}	 	                         & 	0.126	 & 	0.298    & 0.227	 & 	0.017	 & 	0.013	 & 	-0.034	 & 	0.009	 & 	0.579	 & 	0.587	 & 	0.682	 & 	-1.445	 & 	5.421	 & 	0.056        \\
\rowcolor{LightCyan}  {\footnotesize E05,16,9}	    	 & 	0.112	 & 	0.308    & 0.167	 & 	0.018	 & 	0.014	 & 	-0.031	 & 	0.015	 & 	0.637	 & 	0.636	 & 	0.753	 & 	-0.738	 & 	3.714	 & 	0.043       \\
 {\footnotesize E06,16,15}	                             & 	0.081	 & 	0.191    & 0.191    & 	0.013	 & 	0.010	 & 	-0.027	 & 	0.003	 & 	0.578	 & 	0.621	 & 	0.713	 & 	-0.687	 & 	3.854	 & 	0.035        \\
\rowcolor{LightCyan}  {\footnotesize E07,18,1}	    	 & 	0.121	 & 	0.241	 & 0.227    & 	0.026	 & 	0.021	 & 	-0.013	 & 	-0.183	 & 	0.500	 & 	0.181	 & 	0.188	 & 	0.107	 & 	2.941	 & 	0.053      \\
 {\footnotesize E08,18,2}	                        	 & 	0.143	 & 	0.276    & 0.255	 & 	0.032	 & 	0.025	 & 	-0.019	 & 	0.194	 & 	0.483	 & 	0.162	 & 	0.164	 & 	0.093	 & 	2.967	 & 	0.034     \\
\rowcolor{LightCyan}  {\footnotesize E09,19,1}	    	 & 	0.204	 & 	0.398    & 0.344	 & 	0.046	 & 	0.036	 & 	0.042	 & 	-0.096	 & 	0.487	 & 	0.227	 & 	0.230	 & 	-0.080	 & 	2.989	 & 	0.065     \\
 {\footnotesize E10,19,2}		                         & 	0.389	 & 	0.763    & 0.689	 & 	0.088	 & 	0.070	 & 	0.046	 & 	0.002	 & 	0.492	 & 	0.447	 & 	0.452	 & 	-0.065	 & 	2.925	 & 	0.200     \\
\rowcolor{LightCyan}  {\footnotesize E11,19,3}	    	 & 	0.477	 & 	0.730    & 0.679    & 	0.088	 & 	0.070	 & 	-0.029	 & 	-0.245	 & 	0.348	 & 	0.434	 & 	0.432	 & 	-0.660	 & 	3.274	 & 	0.160     \\
 {\footnotesize E12,19,4}	                           	 & 	0.459	 & 	0.751    & 0.710	 & 	0.089	 & 	0.071	 & 	-0.052	 & 	0.036	 & 	0.391	 & 	0.455	 & 	0.459	 & 	-0.351	 & 	3.041	 & 	0.180      \\
\rowcolor{LightCyan}  {\footnotesize E13,19,5}	    	 & 	0.292	 & 	0.732    & 0.650	 & 	0.090	 & 	0.072	 & 	-0.058	 & 	-0.004	 & 	0.602	 & 	0.445	 & 	0.452	 & 	0.346	 & 	3.051	 & 	0.245       \\
 {\footnotesize E14,19,6}	                        	 & 	0.202	 & 	0.711    & 0.604	 & 	0.087	 & 	0.069	 & 	0.004	 & 	-0.010	 & 	0.716	 & 	0.391	 & 	0.400	 & 	0.812	 & 	3.559	 & 	0.435      \\	
\rowcolor{LightCyan}  {\footnotesize E15,19,7}	    	 & 	0.522	 & 	0.967    & 0.894	 & 	0.114	 & 	0.092	 & 	-0.050	 & 	-0.235	 & 	0.462	 & 	0.557	 & 	0.562	 & 	-0.066	 & 	2.794	 & 	0.230    \\  

     \end{tabular}
     \caption{Statistical parameters of roughness topography and the equivalent sand-grain height $k_s$ for each roughness geometry. $R_a$, $k_{avg}$, $k_c$, $k_t$, $k_{rms}$ and $k_s$ values from DNS are normalized by the channel half height $\delta$, while corresponding experimental values are given in \textit{mm}. $k_s$ is not listed for cases thought to be transitionally rough.}
   \label{tab:roughness2}}
   \end{center}
 \end{table}

In figure \ref{fig:vof}, surface plots of the 45 roughness geometries used in these simulations are displayed; their statistical properties are given in table \ref{tab:roughness2}.
Each case name in figure \ref{fig:vof} and table \ref{tab:roughness2} begins with the letter C or E, which denotes whether the data is computational or experimental, followed by an identifying index for that particular surface. For computational cases, this index is followed by: a characteristic length scale (as a percentage of $\delta$) used for roughness synthesis; an identifier of whether the surface roughness is regular (reg) or random (rnd); and finally an identifier for one additional surface feature and its position in a series of surfaces with different sizes of that feature. These features were: the streamwise inclination angle $I_x$ in surfaces C01 to C12; the porosity $P_o$ in surfaces C13 to C24; and the streamwise effective slope $E_x$ in surfaces C25 to C30. 
For the experimental data two  indices were  assigned to each surface. The first denotes the year in which the data were published and the second is the surface designation in that publication. Thus surfaces with index 16 are from  \citet{FlackFSBK16a}, those with index 18 are from  \citet{BarrosBSF18a}, and those with index 19 are from  \citet{FlackFSB19a}. Note that these experimental data were obtained from fully-developed channel flows,  where the drag was measured through the pressure drop along the channel. Thus their results are expected to be more accurate than those of  boundary layer studies where the drag is usually inferred.

Surfaces C01 through C24 were created using ellipsoidal elements \citep{Scotti06} of different size, aspect ratio and inclination. For \textit{regular} roughness, each element had the same orientation and semi-axis lengths, $(\lambda_1,\lambda_2,\lambda_3) = (1.0, 0.7, 0.5)k_c$, where $k_c$ is the peak-to-trough height (also called the crest height). For \textit{random} roughness, the elements had random orientations and semi-axis lengths (with uniform distributions of the random variables). The average orientation and semi-axis lengths for \textit{random} roughness were the same as the corresponding \textit{regular} surface. Surfaces C25 through C30 comprised  sinusoidal waves in the $x$ direction, of the same magnitude but different wavelengths, to generate different values of effective slope $E_x$. The  wavelengths were  $3\delta/4$,~$3\delta/8$ and $\delta/6$. Surfaces C31 and C37 comprised the random sand-grain roughness of Scotti, which were produced by randomly oriented ellipsoidal elements with fixed semi-axes of $(1.0, 0.7, 0.5)k_c$. Surfaces C32 through C36 and C38 through C42 were generated as the low-order (the first 5, 10, 20, 30 and 50) modes of Fourier transforms of white noise in the streamwise and spanwise directions; they therefore describe random surfaces with large- to small-wavelength roughness. Cases C43, C44 and C45 are DNS results from full-span channel computations of flow over surfaces of: random sand-grain roughness; the roughness found on a turbine blade~\citep{YuanA18}; and arrays of cubes  \citep[from the study of][]{Aghaei-JouybariABY19a} respectively. Case C46 is a full-span DNS of case C21, generated to validate the minimal-channel approach of the preceding cases.
A baseline smooth-wall flow was also simulated using a full-span channel \citep{YuanA18}.

The geometric parameters reported for each surface in table~\ref{tab:roughness2} are: roughness peak-to-trough height (also termed crest height) $k_c$ (i.e. distance between the highest  and the lowest surface points); mean peak-to-trough height $k_t$ (i.e. the average  of peak-to-trough heights obtained from surface tiles of size $\delta\times\delta$, similar to \cite{ForooghiFSMJF17a}); mean roughness height $k_{avg}$; first-order moment of height fluctuations $R_a$; root-mean-square $k_{rms}$,  skewness $S_k$ and kurtosis $K_u$ of the roughness height fluctuations; surface porosity $P_o$; effective slope in the $x_i$ direction $E_{x_i}$; and inclination angle (in radians)  in the $x_i$ direction $I_{x_i}$,  together with the hydrodynamic lengthscale $k_s$ deduced from the mean velocity field using equation~(\ref{eq:ks_def}).

These geometrical parameters are defined as:
\begin{equation}
\label{e:kavg}
    k_{avg}  =\frac{1}{A_t}\int_{x,z}kdA,
\end{equation}
\begin{equation}
\label{e:ra}
    R_a        = \frac{1}{A_t}\int_{x,z}\vert k - k_{avg}\vert dA,
\end{equation}
\begin{equation}
\label{e:krms}
    k_{rms}     = \sqrt{\frac{1}{A_t}\int_{x,z} (k-k_{avg})^2 dA} ,
\end{equation}
\begin{equation}
\label{e:S_k}
    S_k        = \left.\frac{1}{A_t}\int_{x,z}(k - k_{avg})^3dA \right/ k_{rms}^3,
\end{equation}
\begin{equation}
\label{e:kur}
    K_u         = \left.\frac{1}{A_t}\int_{x,z}(k - k_{avg})^4dA\right/ k_{rms}^4,
\end{equation}
\begin{equation}
\label{e:ESx}
     E_x       = \frac{1}{A_t}\int_{x,z}\Big|\frac{\partial k}{\partial x} \Big|dA,
\end{equation}
\begin{equation}
\label{e:ESz}
    E_z       = \frac{1}{A_t}\int_{x,z}\Big|\frac{\partial k}{\partial z} \Big|dA,
\end{equation}
\begin{equation}
\label{e:por}
    P_o        = \frac{1}{A_t k_c}\int_0^{k_c} A_f dy,
\end{equation}
\begin{equation}
\label{e:incx}
    I_x    = tan^{-1}\bigg\{\frac{1}{2}S_k\bigg(\frac{\partial k}{\partial x}\bigg)\bigg\},
\end{equation}
\begin{equation}
\label{e:incz}
    I_z    = tan^{-1}\bigg\{\frac{1}{2}S_k\bigg(\frac{\partial k}{\partial z}\bigg)\bigg\},
\end{equation}
where $k(x,z)$ is the roughness height distribution and $A_f(y)$ and  $A_t(y)$ are the fluid and total planar areas at each $y$ location. $S_k(\partial k/\partial x_i)$ is the skewness of $\partial k/\partial x_i$ distribution.   In table~\ref{tab:roughness2},
$k_{avg}$, $k_c$, $k_{rms}$ and $k_s$  are then normalized by the first-order moment of height fluctuations $R_a$ and were incorporated in the ML algorithms in this form. 
All surfaces considered were in the ranges $k_c/\delta\le 0.17$ and $R_a/\delta \le 0.04$.

\subsection{Simulation parameters}
Direct numerical simulation was used to calculate the  velocity and pressure fields in turbulent  open-channel   flows over 45 different rough surfaces and one smooth one, at a constant frictional Reynolds number $\textup{Re}_\tau=u_\tau \delta/\nu = 1000$, where $u_\tau$ is the friction velocity and $\delta$ is the channel half-height. In these simulations, the domain sizes were $(L_x, L_y , L_z)=(3, 1, 1)\delta$.  The origin of the $y$ axis was the elevation of the lowest trough for each rough surface. The number of grid points   was $(n_x, n_y , n_z)=(400, 300, 160)$. A uniform mesh was used in the $x$ and $z$ directions, yielding grid sizes  of $\Delta x^+=7.5$ and $\Delta z^+=6.3$, where ${+}$ denotes normalization in wall units. For all cases, the mesh was stretched in the $y$ direction with a hyperbolic tangent function, with the third grid point from the origin at $y^+ <  1$. For the rough-wall cases, at the roughness crest, $\Delta y/k_c\le 0.017$, with this ratio taking its highest value for Case C11.  The maximum grid size was $\Delta y^+_{\max}=9.5$ at the channel center line, where the Kolmogorov length scale $\eta^+\approx 6$.
\citet{MoinM98} have proposed that one requirement for obtaining reliable first- and second-order flow statistics is that the grid resolution be fine enough to capture accurately most of the dissipation, while   \citet{MoserM87} noted that most of the dissipation in curved channel flow occurs at scales greater than $15\eta$ (based on average dissipation).  It follows that for DNS computations of these kinds of flow statistics in channel and boundary-layer flows, $\Delta x/\eta$ and $\Delta z/\eta$  are typically chosen between 7 to 15, and 4 to 8 respectively (see, for example,  \citet{KimMM87}, \citet{Spalart88} and  \citet{YuanP14b}). The grid sizes in this study were chosen accordingly and were: $\Delta x/\eta<7.5$, $\Delta y/\eta<4.0$, and $\Delta z/\eta<6.5$.

Periodic boundary conditions were imposed in the streamwise and spanwise directions, with no-slip and symmetry boundary conditions at the bottom and top boundaries respectively. After each simulation had reached statistical stationarity, data were collected for ensemble averaging over 10 large-eddy turn-over times ($\delta/u_\tau$). In these simulations, the time step $\tau^+\le 0.04$ and so was significantly smaller than the largest acceptable one of $\tau^+\approx 0.2$  recommended by \citet{ChoiM94} for DNS.

The surface Taylor micro-scales $\lambda_{T,x}$ and $\lambda_{T,z}$, in the $x$ and $z$ directions, were used to evaluate the adequacy of the grid resolution for resolving details of flow in the roughness sublayer, following  \citet{YuanP14}. These geometric micro-scales were obtained by fitting a parabola to the two-point autocorrelation of the surface height fluctuation in the respective direction. They represent the size of an equivalent `roughness element' in the context of random multiscale roughness.  The streamwise and spanwise values of $\lambda_T$, rescaled by $u_\tau/\nu$ as $\lambda_T^+$, and the respective grid sizes are given in table~\ref{tab:lamb_r} (part I). For each case, $\lambda_{T,x_i}^+$ is of order $10$ to $10^2$, indicating that the average size of the roughness element is large in viscous units.  On average, roughness elements  were well resolved by the grid, with typically 4 to 12 grid points  per $\lambda_{T,x_i}$ microscale in each direction. For reference purposes, \citet{YuanP14a} reported a resolution of $\lambda_{T,x}/\Delta x\approx 4$ in their large-eddy simulations of channel flow over surfaces with sand-grain roughness. The cases in table~\ref{tab:lamb_r} for which $\lambda_T$ was not well resolved in at least one direction ($\lambda_{T,x}/\Delta x < 3$ or $\lambda_{T,z}/\Delta z <3$) may also not have been fully-rough flows (as discussed in the following section), and so were not included in the ensemble of flows for ML training and testing. 

\begin{table}
  \begin{center}{\footnotesize 
    \def~{\hphantom{10}}
    \begin{tabular}{l c c c c c c c}
        &  \multicolumn{4}{c}{Part I} & & \multicolumn{2}{c}{Part II}      \\    \cline{2-5} \cline{7-8} \\
 Case name	 &   $\lambda_{T,x}^+$	 & 	$\lambda_{T,x}/\Delta x$ &   $\lambda_{T,z}^+$	 & 	$\lambda_{T,z}/\Delta z$ && $d/\delta$  & $\widehat{k}_s^+$   \\ 
C01,r4,reg,inc1	 & 	19.7	 & 	2.6	 & 	21.1	 & 	3.4	 && 	0.032	 & 	19.4	 \\ 
C02,r4,reg,inc2	 & 	20.4	 & 	2.7	 & 	33.1	 & 	5.3	 && 	0.046	 & 	49.7	 \\ 
C03,r4,reg,inc3	 & 	19.8	 & 	2.6	 & 	22.9	 & 	3.7	 && 	0.033	 & 	31.0	 \\ 
C04,r6,reg,inc1	 & 	27.7	 & 	3.7	 & 	28.4	 & 	4.5	 && 	0.038	 & 	64.4	 \\ 
C05,r6,reg,inc2	 & 	31.6	 & 	4.2	 & 	39.1	 & 	6.2	 && 	0.057	 & 	124.4	 \\ 
C06,r6,reg,inc3	 & 	29.9	 & 	4.0	 & 	30.0	 & 	4.8	 && 	0.045	 & 	58.9	 \\ 
C07,r4,rnd,inc1	 & 	33.8	 & 	4.5	 & 	26.7	 & 	4.3	 && 	0.036	 & 	136.2	 \\ 
C08,r4,rnd,inc2	 & 	26.1	 & 	3.5	 & 	32.7	 & 	5.2	 && 	0.052	 & 	322.3	 \\ 
C09,r4,rnd,inc3	 & 	35.5	 & 	4.7	 & 	30.1	 & 	4.8	 && 	0.039	 & 	131.1	 \\ 
C10,r6,rnd,inc1	 & 	38.2	 & 	5.1	 & 	29.7	 & 	4.8	 && 	0.042	 & 	268.9	 \\ 
C11,r6,rnd,inc2	 & 	38.1	 & 	5.1	 & 	47.0	 & 	7.5	 && 	0.070	 & 	536.4	 \\ 
C12,r6,rnd,inc3	 & 	47.9	 & 	6.4	 & 	40.2	 & 	6.4	 && 	0.053	 & 	271.7	 \\ 
C13,r4,reg,por1	 & 	17.8	 & 	2.4	 & 	32.7	 & 	5.2	 && 	0.047	 & 	41.4	 \\ 
C14,r4,reg,por2	 & 	27.5	 & 	3.7	 & 	34.2	 & 	5.5	 && 	0.032	 & 	140.6	 \\ 
C15,r4,reg,por3	 & 	31.5	 & 	4.2	 & 	39.4	 & 	6.3	 && 	0.028	 & 	157.1	 \\ 
C16,r6,reg,por1	 & 	25.6	 & 	3.4	 & 	46.1	 & 	7.4	 && 	0.066	 & 	76.7	 \\ 
C17,r6,reg,por2	 & 	40.1	 & 	5.3	 & 	47.8	 & 	7.6	 && 	0.044	 & 	259.8	 \\ 
C18,r6,reg,por3	 & 	44.4	 & 	5.9	 & 	54.8	 & 	8.8	 && 	0.039	 & 	246.5	 \\ 
C19,r4,rnd,por1	 & 	32.7	 & 	4.4	 & 	31.1	 & 	5.0	 && 	0.042	 & 	158.2	 \\ 
C20,r4,rnd,por2	 & 	35.6	 & 	4.7	 & 	31.3	 & 	5.0	 && 	0.026	 & 	105.7	 \\ 
C21,r4,rnd,por3	 & 	37.4	 & 	5.0	 & 	34.2	 & 	5.5	 && 	0.027	 & 	102.7	 \\ 
C22,r6,rnd,por1	 & 	44.6	 & 	5.9	 & 	35.3	 & 	5.6	 && 	0.053	 & 	276.8	 \\ 
C23,r6,rnd,por2	 & 	47.1	 & 	6.3	 & 	39.7	 & 	6.4	 && 	0.038	 & 	175.1	 \\ 
C24,r6,rnd,por3	 & 	47.1	 & 	6.3	 & 	44.4	 & 	7.1	 && 	0.045	 & 	260.3	 \\ 
C25,r4,reg,ES1	 & 	89.0	 & 	11.9	 & 	--	 & 	--	 && 	0.024	 & 	25.6	 \\ 
C26,r4,reg,ES2	 & 	66.5	 & 	8.9	 & 	--	 & 	--	 && 	0.026	 & 	65.3	 \\ 
C27,r4,reg,ES3	 & 	27.1	 & 	3.6	 & 	--	 & 	--	 && 	0.035	 & 	45.5	 \\ 
C28,r6,reg,ES1	 & 	90.6	 & 	12.1	 & 	--	 & 	--	 && 	0.033	 & 	71.2	 \\ 
C29,r6,reg,ES2	 & 	66.8	 & 	8.9	 & 	--	 & --	 && 	0.040	 & 	112.0	 \\ 
C30,r6,reg,ES3	 & 	27.2	 & 	3.6	 & 	--	 & 	--	 && 	0.054	 & 	64.0	 \\ 
C31,r4,rnd,SGR	 & 	27.8	 & 	3.7	 & 	25.0	 & 	4.0	 && 	0.032	 & 	48.7	 \\ 
C32,r4,rnd,RND1	 & 	131.2	 & 	17.5	 & 	54.1	 & 	8.7	 && 	0.041	 & 	8.4	 \\ 
C33,r4,rnd,RND2	 & 	96.3	 & 	12.8	 & 	42.1	 & 	6.7	 && 	0.043	 & 	17.6	 \\ 
C34,r4,rnd,RND3	 & 	56.4	 & 	7.5	 & 	22.4	 & 	3.6	 && 	0.045	 & 	22.5	 \\ 
C35,r4,rnd,RND4	 & 	39.5	 & 	5.3	 & 	15.8	 & 	2.5	 && 	0.046	 & 	18.3	 \\ 
C36,r4,rnd,RND5	 & 	25.1	 & 	3.3	 & 	11.4	 & 	1.8	 && 	0.051	 & 	23.4	 \\ 
C37,r6,rnd,SGR	 & 	36.5	 & 	4.9	 & 	31.9	 & 	5.1	 && 	0.046	 & 	108.8	 \\ 
C38,r6,rnd,RND1	 & 	88.5	 & 	11.8	 & 	72.6	 & 	11.6	 && 	0.060	 & 	12.0	 \\ 
C39,r6,rnd,RND2	 & 	93.8	 & 	12.5	 & 	35.7	 & 	5.7	 && 	0.062	 & 	17.1	 \\ 
C40,r6,rnd,RND3	 & 	57.0	 & 	7.6	 & 	22.8	 & 	3.6	 && 	0.070	 & 	50.4	 \\ 
C41,r6,rnd,RND4	 & 	40.5	 & 	5.4	 & 	15.6	 & 	2.5	 && 	0.073	 & 	48.7	 \\ 
C42,r6,rnd,RND5	 & 	24.5	 & 	3.3	 & 	11.3	 & 	1.8	 && 	0.076	 & 	43.8	 \\ 
C43,SG	 & 	35.2	 & 	6.0	 & 	33.5	 & 	5.7	 && 	0.044	 & 	93.0	 \\ 
C44,TB	 & 	132.1	 & 	10.4	 & 	168.5	 & 	13.2	 && 	0.058	 & 	24.1	 \\ 
C45,CB	 & 	25.7	 & 	4.5	 & 	25.5	 & 	4.4	 && 	0.039	 & 	149.9	 \\ 
C46,r4,rnd,por3,FS	 & 	37.6	 & 	5.0	 & 	34.6	 & 	5.5	 && 	0.027	 & 	104.2	 \\ 
    \end{tabular}
    \caption{Part I: Streamwise and spanwise values of the surface Taylor micro-scale $\lambda_T$. Part II: Flow-related parameters obtained from DNS. The flow is assumed fully rough if $\widehat{k}_s^+\gtrsim 50$, in which case $k_s$ is equal to $\widehat{k}_s$.}
   \label{tab:lamb_r}}
  \end{center}
\end{table}

In rough-wall flows, the pressure drag is caused primarily by the local flow structures  and separation in the vicinity of individual roughness protuberances, which are predominately near-wall phenomena. To carry out the 46 separate DNS simulations for determining $k_s$ efficiently, with sufficient near-wall resolution, a small-span channel simulation approach was employed. The concept of minimal-span simulation was introduced by  \citet{JimenezJM91a}.  \citet{ChungCCMHO15} and  \citet{MacDonaldMCHCO17} carried out analyses of the performance of DNS over small spanwise domains for full and open channel flows on rough and smooth walls and showed that minimal-span simulations captured the essential near-wall dynamics and yielded accurate computations of wall friction, and of mean velocities and Reynolds stresses as far from the wall as $y\approx 0.3\delta$, when the following constraints were met:
\begin{subequations}
\begin{align}
     L_x   & \ge  \max\big(1000 \delta_\nu,3L_z,\lambda_{r,x}\big),\\
     L_y   & \ge  k_c/0.15, \\
     L_z   & \ge  \max\big(100\delta_\nu,k_c/0.4,\lambda_{r,z}\big),
\end{align}\label{eq:min_chn}
\end{subequations}
where $\delta_\nu=\nu/u_\tau$ and $\lambda_{r,x_i}$ is the characteristic roughness wavelength in the $x_i$ direction. Alternatively,  the  surface  Taylor  microscale  may  be used  as  the  lengthscale  in  this constraint.
Conditions (\ref{eq:min_chn}\textit{a,c}) were satisfied by choosing domain sizes $L_x^+$ and $L_z^+$ of 3000 and 1000 respectively, while condition (\ref{eq:min_chn}\textit{b}) was met for all cases except C11, which fell below the $L_y\ge k_c/0.15$ constraint by about 10\%. 
C11 is a case with random geometry; protuberances beyond $0.15\delta$ exist but are rare.

\begin{figure}
   \centerline{\includegraphics[width=1\textwidth,trim={0 0cm 0 0cm},clip]{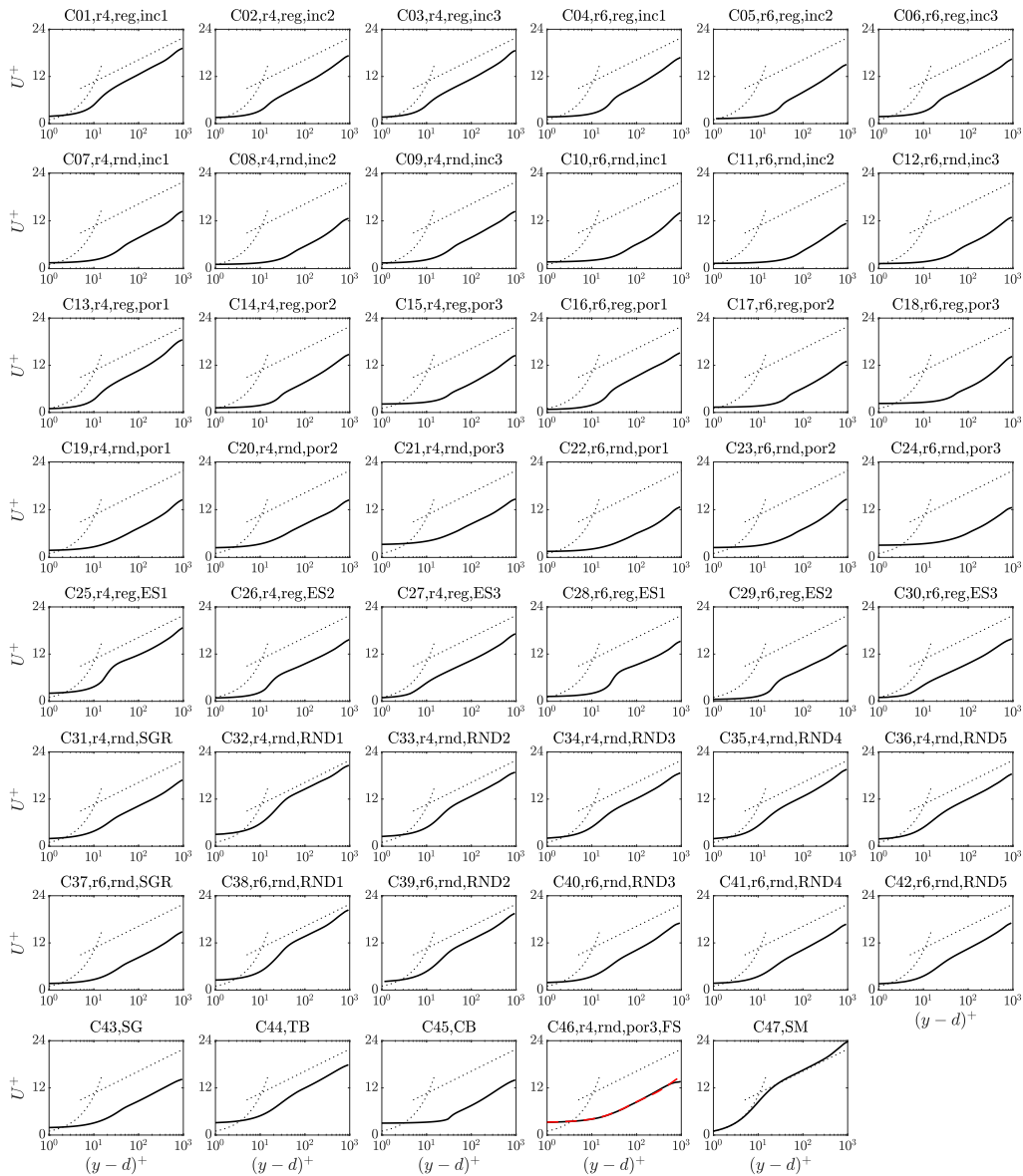}}
   \caption{Profiles of streamwise double-averaged velocity plotted against a zero-plane-displacement shifted logarithmic $y$ abscissa. The dashed lines are $u^+=y^+$ and $u^+=2.5\ln{(y-d)^+}+5.0$. The red dot-dash line in plot C46 is that of C21. }
\label{fig:upyp2}
\end{figure}

The criteria of (\ref{eq:min_chn}) were developed originally for simulations of flow over surfaces with uniformly distributed roughness elements. In this study, the random roughness geometries used require an additional criterion on the sufficiency of the domain size: the area $L_xL_z$ should be large enough to achieve statistical convergence of surface parameters, such as $k_{rms}$ and $E_{x_i}$, and of the flow parameter $k_s$. To check the adequacy of the chosen domain size, an additional simulation was carried out of Case C21, the surface  comprising the largest dominant spatial wavelength (and consequently the most limited  sampling of random geometrical components  with this wavelength) and a long-tailed height-fluctuation pdf with a kurtosis of around 8. In this validation simulation, denoted Case C46, the domain sizes were doubled in $x$ and $z$, by duplicating C21 in these directions.
The double-averaged velocity profiles $U^+=\langle \overline{u}\rangle^+(y^+)$  for Cases C21 and C46 are in a very good agreement over the log-linear region, as shown in figure \ref{fig:upyp2}. Each surface statistic differs by no more than 3\%, with the greatest discrepancy found  in $I_z$, while the equivalent sandgrain roughness height $k_s$ is almost equal in the two cases. The chosen domain size was therefore considered sufficient for accuracy and convergence of statistics describing flow over the random roughness geometries of this study.

\begin{figure}\centering
    \includegraphics[width=1\textwidth,trim={0 0cm 2cm 0cm},clip]{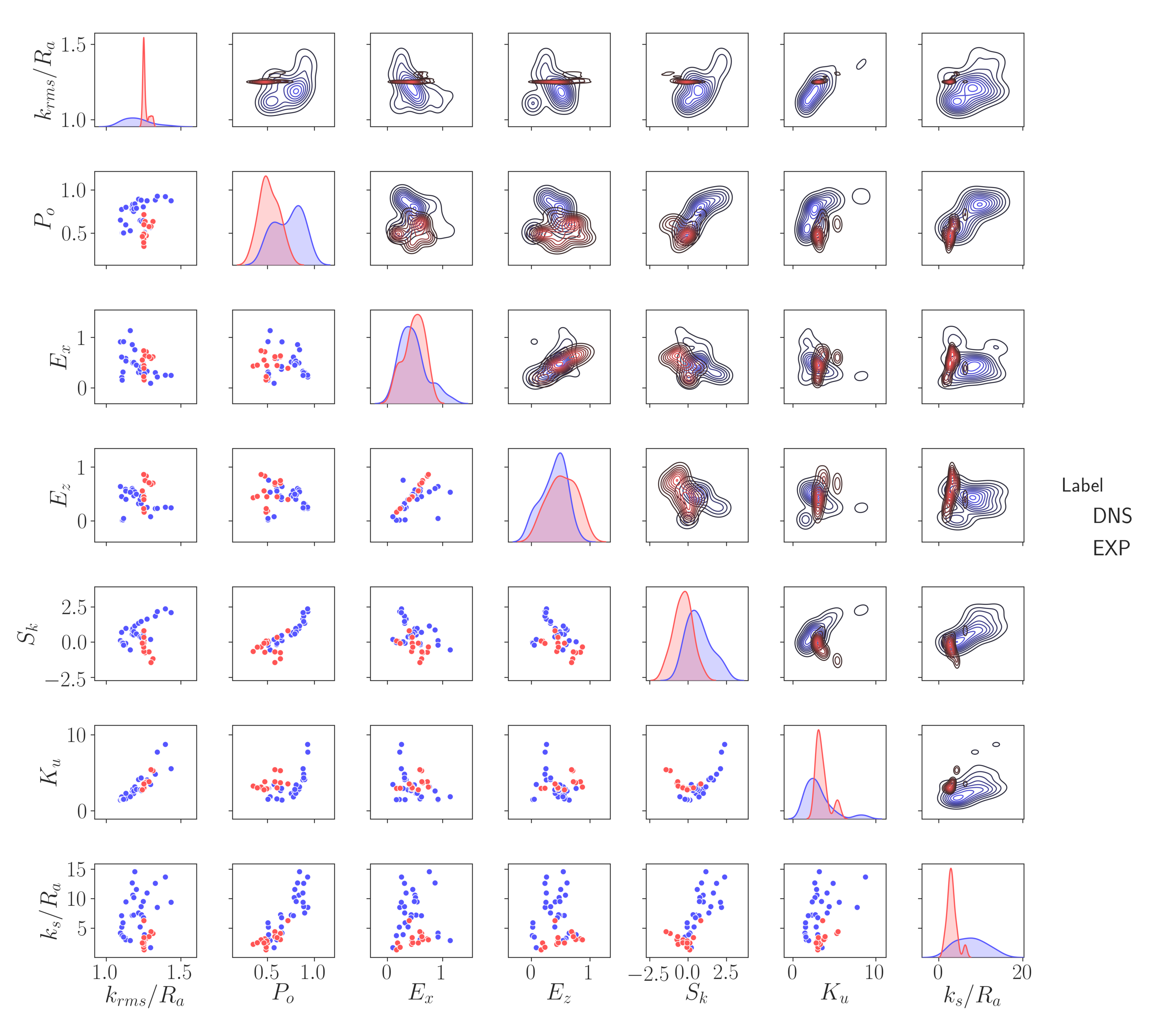}
    \caption{Pair plots of geometrical parameters and $k_s$, with $k_s$ plots in the bottom row and the first column, DNS data (\textit{blue}), experimental data (\textit{red}).}
    \label{fig:pair}
\end{figure}

\section{Results}\label{sec:results}
\subsection{Post-processed results}
In figure \ref{fig:upyp2}, the streamwise double-averaged velocity profiles computed in these simulations are shown. The profiles in the logarithmic region are described for the smooth-wall case and the fully-rough rough-wall cases  as
\begin{subequations}
\begin{align}
     \big<\overline{u}\big>^+ & =\frac{1}{\kappa}\ln(y^+)+5.0   ,\ \hbox{and}\\
     \big<\overline{u}\big>^+ & =\frac{1}{\kappa}\ln\bigg(\frac{y-d}{k_s}\bigg)+8.5
\end{align}\label{eq:law}
\end{subequations}
respectively, where $d$ is the zero-plane displacement, obtained as the location of the centroid of the wall-normal profile of the averaged drag force  \citep{Jackson81}. The shift in the $y$ coordinate by $d$ accounts for the flow blockage by surface roughness elements, and
the values of $d$ are given in table \ref{tab:lamb_r} (part II).

To determine whether a particular flow was within the fully rough regime, equation~(\ref{eq:law}b) was applied to the computed logarithmic velocity profile to yield a test value of $k_s$, denoted as $\widehat{k}_s$ in table \ref{tab:lamb_r} (part II). 
With $\widehat{k}_s$ determined for all cases, those with $\widehat{k}_s^+$ greater than a threshold value of 50 were deemed to be in the fully rough regime (30 surfaces), in which case $k_s$ was set to equal $\widehat{k}_s$. Those below the threshold were possibly transitionally rough (15 surfaces) and so were not included in ML predictions in this study. The threshold value of  $k_s^+$---the lower end of the fully rough regime---has been observed to vary significantly for different types of roughness and is typically between 20 and 80. For example, the threshold values for surfaces C43 and C44 are roughly 80 and 20 \citep{YuanP14a}, and 50 for surface C45 \citep{Bandyopadhay87}.

The threshold value of $k_s^+$ which signifies the beginning of the fully rough regime was not determined more precisely because of the cost of carrying out, for each surface, simulations at successively higher values of $k_s^+$ until $k_s/R_a$ became invariant with the Reynolds number. In the GPR prediction, potential uncertainties in $k_s$ which might arise through treating all flows with $k_s^+>50$ as fully rough, and other sources of possible error, were compensated for by incorporating an assumed 10 \%  noise level in the learning stage of the prediction of $k_s$, as discussed in section \ref{sec:ML}.
The values of $k_s^+=50$ as the threshold for fully rough flows and the assumed noise level were chosen as part of a trade-off to maximize the number of usable data, to avoid overfitting, while acknowledging possible uncertainties in the modeling data.

In figure \ref{fig:pair}, pair plots of the different topographic roughness parameters are shown as scatter plots (lower left), joint pdfs (upper right), and distribution pdfs (diagonal).
Pair scatter plots for the true (DNS and experimental) value of $k_s$ 
and other roughness parameters are along the bottom row of this figure. It can be seen that, for the roughness cases chosen, there is some correlation between kurtosis and rms roughness (column 1, row 6), kurtosis and skewness (column 5, row 6), and skewness and porosity (column 2, row 5). The relationship between others appears to be more random. From the graphs in the
bottom row, it can be seen that $k_s/R_a$ scales on porosity to some power, albeit with some scatter (column 2, row 7).
It also appears that $k_s/R_a$ might decrease with skewness for surfaces with $S_k<0$ and increase with skewness in cases with $S_k>0$  (column 5, row 7). Surfaces with positive skewness yielded higher values of $k_s$ compared to those with negative skewness, consistent with the observation of \citet{FlackFSB19a}.
Beyond these observations, there does not appear to be a clear linear correlation between $k_s$ and any individual roughness parameter, which makes the search for 
a functional dependence of $k_s$ on these parameters a problem well suited
to ML. The measures of inclination, $I_x$ and $I_z$, showed no clear correlation with other variables or with $k_s/Ra$.

\subsection{ML predictions of the equivalent sand-grain height}\label{sec:ML}

The ML techniques of DNN and GPR were employed to predict $k_s$ from the data sets described in the previous section.
The objectives of this exercise were to generate and collect data, and make qualitative comparisons between ML predictions and those from conventional correlations, rather than evaluating and comparing the performance of various ML procedures per se.
DNN and GPR approaches were used because our experience was that they predicted $k_s$ with high accuracy, notwithstanding their simplicity. Other approaches such as the Support Vector Machine (SVM) technique were considered initially, but their preliminary predictions were not as accurate as those found using DNN and GPR approaches.

The main characteristics of DNN and GPR methods are described below:
 \begin{itemize}
 
 \item The inputs for both techniques were 17 roughness geometrical parameters, 8 of which were the primary variables $k_{rms}/R_a$, $I_x$, $\vert I_z \vert$, $P_o$, $E_x$, $E_z$, $S_k$ and $K_u$ (defined in equations \ref{e:kavg} to \ref{e:incz}). The other 9 were products of the primary variables, which were added to improve the efficiency of each learning stage. They were $p_1=E_xEz$, $p_2=E_xS_k$, $p_3=E_xK_u$, $p_4=E_zS_k$, $p_5=E_zK_u$, $p_6=S_kK_u$, $p_7=E_x^2$, $p_8=E_z^2$ and $p_9=S_k^2$. These particular products were chosen because of their perceived importance for certain types of roughness.
 \item The database consisted of 45 different sets: 30 DNS of turbulent channel flows over different surfaces at $\hbox{Re}_\tau=1000$, and 15 experimental data sets at higher Reynolds numbers, with all data sets in the fully-rough turbulent-flow regime.

\item The DNN architecture was a \textit{Multi Layer Perceptron}, with three hidden layers (with 18, 7 and 7 neurons respectively). 
The activation functions at all nodes were of the \textit{Rectified Linear Unit} kind, and kernel regularization was used to avoid overfitting. The network had 521 trainable weights in total. 
The preset parameters to the algorithm were optimized based on available data, through a \textit{hyper-parameter tuning} process. Specifically, 270 configurations were first generated with different lengths (representing the number of layers) and widths (representing the  number of neurons). For each configuration, the DNN compiler was performed 1000 times with random selections of training (70\% of total) and testing (30\% of total) datasets to identify the best performance of the configuration.  The configuration   that yielded the best results was considered as the optimal one, the results of which are presented here. The cost of  data fitting for one iteration (out of 1000)  for each DNN configuration  was about one second. In total, it took about 75 hours to obtain the optimal DNN network.
This architecture was found to provide suitable accuracy in modeling without overfitting, for this particular multivariate labeled regression problem.

\item The GPR procedure used \textit{Rational Quadratic} kernels to represent $k_s$ as a superposition of scaled Gaussian functions of the independent variables of the modeling problem.  Similar to the DNN method, the training and testing data were chosen randomly, with respective ratios of 70\% and 30\% of the total data points.   
The preset parameters (e.g. kernel type, number of iterations, etc.) were also tuned with the available data by running the GPR compiler for about 8000 times. It took about 35 hours to obtain the optimal fit.
The GPR method has the capability of incorporating uncertainty or noise in the determination of model parameters in the learning stages. Such noise might arise through: 
numerical and discretization errors; uncertainty in the form and model coefficients of equation (\ref{eq:ks_def}); the applicability and fitting range of equation (\ref{eq:ks_def}) (which was deduced from high Reynolds number experiments) to simulations at much lower 
Reynolds numbers; and the possibility that some of the training data may have been from simulations in which the flow was not 
quite fully rough. A noise level of 10\% in $k_s/R_a$ values was chosen as an upper estimate of the likely uncertainty 
from these sources. Noise levels of 5\% and 15\% were also tested, but little sensitivity of the $k_s$ prediction was found to the assumed noise level within the tested range.

 \end{itemize}

\begin{figure}\centering
    \includegraphics[width=.8\textwidth]{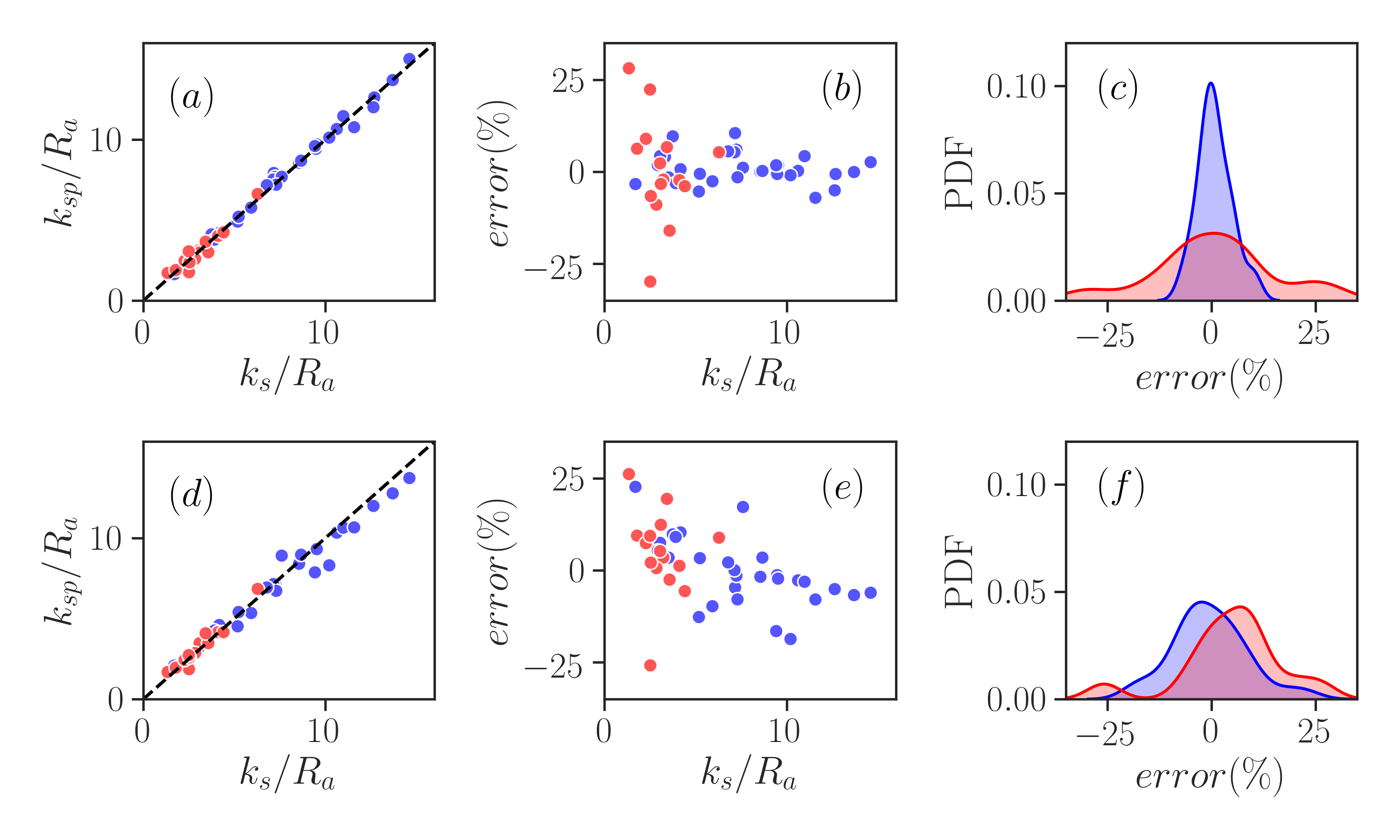}
    \caption{ (a,d) Scatter plot of true $k_s$ and predicted $k_s$, (b,e) scatter plot of true $k_s$ and relative error, (c,f) pdfs of relative error for (a-c) DNN and (d-f) GPR predictions, with DNS data (\textit{blue}),  experimental data (\textit{red}). }
    \label{fig:err_both}
\end{figure}

The   values of $k_s$ predicted from the surface topography parameters, henceforth called $k_{sp}$, are compared to the actual $k_s$ values in figure \ref{fig:err_both}, for the DNN and GPR methods respectively. Scatter plots of  $k_{sp}$  and the true value of $k_s$ in figures \ref{fig:err_both}(a) and (d)  reveal a tight clustering of data along the $y=x$ diagonal, with only a few outlying points. This very high degree of correlation between $k_{sp}$ and $k_s$ implies that both techniques have been applied with equal success to this prediction problem. The error range, figures \ref{fig:err_both}(b) and (e), is less than $\pm 30$\% ($L_\infty$ norm) and the average error ($L_1$ norm) is less than 8\%, for both techniques. 

The consistency between both the $k_s$ predictions and error bands for two quite different ML techniques suggests that they are both well-suited to this kind of problem, and possibly close to an optimum for this class of ML approach.

The error values as percentages, for the DNN and GPR methods, are given in table \ref{tab:err}, together with the error in the empirical relation
\begin{equation}
\label{e:pred_KS}
    k_s=2.91k_{rms}(2+S_k)^{-0.284},
\end{equation}
proposed by  \citet{FlackFSBK16a}, and  \
\begin{equation}
\label{e:pred_KF}
    k_s=1.07k_{t}(1-e^{-3.5 E_x})(0.67 S_k^2+0.93 S_k +1.3),
\end{equation}
given by  \citet{ForooghiFSMJF17a}, as well as their respective recalibrated correlations:
\begin{equation}
\label{e:pred_KS1}
    k_s=1.11k_{rms}(2+ S_k)^{0.74},
\end{equation}
\begin{equation}
\label{e:pred_KF1}
    k_s=0.04k_{t}(1-e^{-5.50 E_x})(S_k^2+2.57 S_k +9.82).
\end{equation}
when extended to all cases in the current database. 
It is interesting to note that the form of equation~(\ref{e:pred_KS}) was chosen for surfaces generated by grit blasting---closely-packed, random, three-dimensional roughnesses with a wide range of scales (E01-E06), while many of the simulated surfaces are two-dimensional, some are characterized by discrete elements of similar sizes, while others are sparse or wavy (characterized by low slopes). Equation (\ref{e:pred_KF}), on the other hand, includes a slope parameter and was calibrated for numerically generated surfaces consisting elements of random sizes and a prescribed shape.

\begin{table}
  \begin{center}{\small 
    \def~{\hphantom{10}}
    \begin{tabular}{l c c c c c c}
Case name &    $err_{DNN}$	 & 	$err_{GPR}$	 & 	$err_{B1}$  & 	$err_{B2}$  & 	$err_{B3}$  & 	$err_{B4}$	 \\ 
C04,r6,reg,inc1	 &4.0	 &4.1	 &-16.7	 &-40.9	 &8.6	 &-63.9\\
C05,r6,reg,inc2	 &0.7	 &10.3	 &-38.3	 &-49.5	 &2.0	 &-71.7\\
C06,r6,reg,inc3	 &4.2	 &7.5	 &-10.4	 &-33.6	 &24.5	 &-59.8\\
C07,r4,rnd,inc1	 &10.5	 &-4.7	 &-63.5	 &-63.6	 &10.0	 &-73.5\\
C08,r4,rnd,inc2	 &-0.6	 &-4.8	 &\q{-80.1}	 &\q{-77.6}	 &-4.1	 &-81.7\\
C09,r4,rnd,inc3	 &6.0	 &-1.5	 &-63.4	 &-64.2	 &8.3	 &-73.4\\
C10,r6,rnd,inc1	 &0.2	 &-2.7	 &-76.3	 &-72.5	 &11.8	 &-77.8\\
C11,r6,rnd,inc2	 &2.6	 &-6.1	 &\q{-82.9}	 &\q{-78.9}	 &4.1	 &\q{-82.2}\\
C12,r6,rnd,inc3	 &-1.0	 &-18.7	 &-74.7	 &-72.7	 &-2.3	 &-78.7\\
C14,r4,reg,por2	 &5.3	 &0.0	 &-66.2	 &-64.2	 &-8.7	 &-80.3\\
C15,r4,reg,por3	 &4.2	 &-3.1	 &-76.7	 &-66.5	 &29.0	 &-78.8\\
C16,r6,reg,por1	 &1.8	 &5.4	 &3.5	 &-41.7	 &21.5	 &-59.4\\
C17,r6,reg,por2	 &-0.6	 &-1.3	 &-74.5	 &-70.3	 &-10.9	 &\q{-82.7}\\
C18,r6,reg,por3	 &-5.1	 &-5.1	 &-79.1	 &-68.4	 &35.5	 &-78.8\\
C19,r4,rnd,por1	 &1.8	 &-2.3	 &-71.4	 &-69.4	 &44.9	 &-67.8\\
C20,r4,rnd,por2	 &1.1	 &17.2	 &-67.0	 &-56.6	 &82.1	 &-66.3\\
C21,r4,rnd,por3	 &0.0	 &-1.8	 &-69.6	 &-50.0	 &254.1	 &-46.2\\
C22,r6,rnd,por1	 &-7.1	 &-7.9	 &-77.0	 &-76.7	 &-10.6	 &-78.4\\
C23,r6,rnd,por2	 &0.2	 &3.4	 &-70.9	 &-60.4	 &80.8	 &-67.9\\
C24,r6,rnd,por3	 &-0.1	 &-6.7	 &\q{-80.5}	 &-66.3	 &136.7	 &-66.5\\
C26,r4,reg,ES2	 &-5.4	 &-12.7	 &-48.6	 &-61.6	 &-57.6	 &\q{-83.8}\\
C28,r6,reg,ES1	 &9.6	 &9.8	 &-29.2	 &-45.9	 &-51.9	 &-81.2\\
C29,r6,reg,ES2	 &-2.6	 &-9.8	 &-54.7	 &-66.2	 &-53.2	 &\q{-83.2}\\
C30,r6,reg,ES3	 &-1.5	 &3.4	 &-21.8	 &-45.7	 &8.1	 &-65.7\\
C31,r4,rnd,SGR	 &-0.6	 &3.3	 &-46.7	 &-50.7	 &65.1	 &-53.8\\
C37,r6,rnd,SGR	 &-1.5	 &-7.9	 &-61.3	 &-65.0	 &11.9	 &-68.6\\
C40,r6,rnd,RND3	 &-3.1	 &9.1	 &-23.6	 &-39.6	 &98.3	 &-30.8\\
C43,SG	 &5.5	 &2.1	 &-58.6	 &-60.1	 &46.3	 &-62.0\\
C44,TB	 &-3.3	 &\q{22.7}	 &77.6	 &51.9	 &31.5	 &-51.6\\
C45,CB	 &1.8	 &-16.5	 &-70.4	 &-52.0	 &79.3	 &-72.8\\
E01,16,2	 &-2.1	 &3.5	 &6.2	 &-47.5	 &370.2	 &63.0\\
E02,16,3	 &2.3	 &5.2	 &3.3	 &-33.7	 &\q{429.4}	 &79.5\\
E03,16,7	 &-2.3	 &1.2	 &-2.2	 &-69.1	 &368.1	 &38.6\\
E04,16,8	 &-3.9	 &-5.7	 &1.3	 &\q{-78.8}	 &\q{412.4}	 &27.6\\
E05,16,9	 &-3.3	 &12.4	 &10.9	 &-46.3	 &262.1	 &27.3\\
E06,16,15	 &\q{-16.0}	 &-2.5	 &-3.0	 &-51.1	 &\q{405.4}	 &79.9\\
E07,18,1	 &\q{-29.8}	 &\q{-25.8}	 &17.3	 &-4.0	 &208.3	 &11.2\\
E08,18,2	 &\q{28.1}	 &\q{26.1}	 &\q{120.7}	 &\q{79.4}	 &\q{388.8}	 &80.0\\
E09,19,1	 &6.2	 &9.4	 &69.2	 &25.9	 &312.5	 &56.9\\
E10,19,2	 &-8.9	 &0.6	 &5.8	 &-20.7	 &258.9	 &20.6\\
E11,19,3	 &8.9	 &7.4	 &47.4	 &-24.1	 &247.4	 &32.2\\
E12,19,4	 &-6.6	 &2.1	 &24.1	 &-21.0	 &258.4	 &32.2\\
E13,19,5	 &6.7	 &\q{19.4}	 &-16.6	 &-23.8	 &287.2	 &6.6\\
E14,19,6	 &5.3	 &8.9	 &-56.8	 &-52.5	 &177.2	 &-38.2\\
E15,19,7	 &\q{22.3}	 &9.4	 &19.8	 &-10.2	 &342.6	 &43.0\\\hline
$L_1$    & 5.4 & 7.8 & 47.6 & 52.8 & 133.8 & 60.6\\
$L_\infty$& 29.8 & 26.1 & 120.7 & 79.4 & 429.4 &83.8\\\hline\hline

    \end{tabular}
     \caption{Errors in $k_s$ prediction by DNN and GPR compared to  errors of the  empirical correlations: $err_{B1}$ (equation \ref{e:pred_KS}), $err_{B2}$ (equation \ref{e:pred_KS1}), $err_{B3}$ (equation \ref{e:pred_KF}) and $err_{B4}$ (equation \ref{e:pred_KF1}). The four largest errors (in magnitude) for each column are colored in red.  The errors are percentages. }
     \label{tab:err}}
  \end{center}
\end{table}

For most cases, the errors from the DNN and GPR methods were of the same  order of magnitude and much smaller than the error in using equation~(\ref{e:pred_KS}) or (\ref{e:pred_KF}). 
In the DNN and GPR predictions of simulation cases, the greatest errors (about 25\%-30\%) arose in cases E07 and E08. The surfaces associated with these cases are characterized by fractal features (with spectral slopes of -0.5 and -1.0, respectively \citep{BarrosBSF18a}). The size of the errors for these cases might be attributed to the small number of surfaces with this feature used in the training set (as opposed to the many surfaces that are mostly characterized by single-scale elements).
A close examination of the prediction errors for the DNS cases  showed a subtle trend between relatively high errors  and low roughness solidity (or low $E_s$ and insignificant wake sheltering), in, for example, cases C28 and C44. Both these cases are characterized by  large-wavelength, wavy features, suggesting an under-representation of sparse roughness in the dataset. Beyond this observation, no   clear correlation was found between the error and other primary roughness parameters included herein  or   surface categorizations (2D/3D, random/regular).

The errors associated with using equation~(\ref{e:pred_KS}) are small for surfaces E01 through E06, which were used to calibrate this relation.  The errors in using equations (\ref{e:pred_KS}) and (\ref{e:pred_KF}) over all surfaces in the database are 120\% and 430\% respectively. However, when recalibrated against the full database, equations (\ref{e:pred_KS1}) and (\ref{e:pred_KF1}) have a significantly smaller error band with maximum values of 79\% and 84\%. The high error values of the empirical correlations, compared to DNN or GPR prediction, are attributed to the small number of geometrical variables used in their calibrations and the restricted range of the models' parameters.

\subsection{Uncertainty estimation}

 \begin{figure}\centering
    \includegraphics[width=1\textwidth,trim={11cm 11cm 10.1cm 14cm},clip]{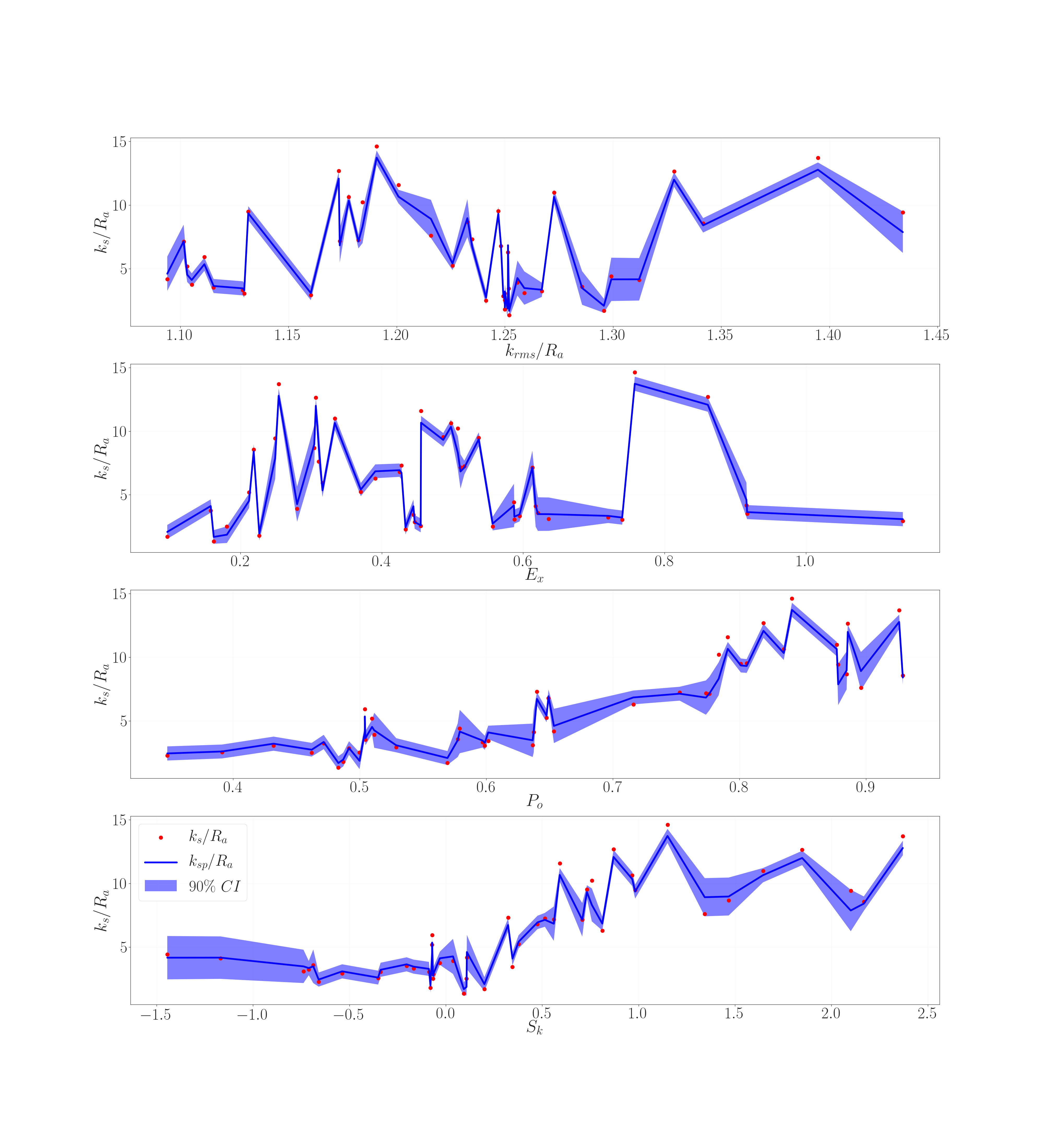}
    \caption{Confidence interval (CI) of predictions with the GPR method, with predicted values of $k_s/R_a$ in \textit{blue lines} (called $k_{sp}$) and true values of $k_s/R_a$ in \textit{red dots}.  GPR predictions for both training and testing data sets are shown --- $k_s$ and $k_{sp}$ are very close to each other for the training data points, while they deviate (less than 30\% of error) for some test data points. Line jaggedness is  associated with  projection of a high-dimensional space to  one-dimensional ones.  }
    \label{fig:gpr}
\end{figure} 

In addition to predictions of equivalent sand-grain height, the GPR method provides confidence margins as functions of each input parameter. These margins can be useful for indicating the kinds of surfaces for which additional training data could improve confidence in predictions. This feature of the GPR approach makes it very attractive for studies of this kind, since DNS and experimental generation of data can be expensive. 

The confidence intervals determined by the GPR technique are shown  as  functions of the normalized surface rms roughness height, effective slope, porosity and skewness in figure \ref{fig:gpr}. Wider intervals indicate higher estimated values of predictive error, such as at roughness porosity of 0.68, and skewnesses of -1.5 and 2.0. Surfaces of roughness with similar values of porosity and skewness would then be priorities for additional simulations or experiments.

\subsection{Sensitivity analysis}  

The dependence of DNN predictions of $k_s$ on individual roughness parameters is explored by determining the change in the error norms when each of the primary surface parameters is removed from the data from which the DNN prediction was made. In table~\ref{tab:err_sens}, the actual error for each surface, and the values of the $L_1$ and $L_\infty$ norms of errors in the prediction of $k_s$ over the 45 surfaces, are reported when the parameter(s) in the first row is (are) the excluded one(s). The errors of the  base prediction (which includes  all 8 primary parameters) are  listed in the second column. In the following discussion, we focus on the $L_1$ norm for ease of comparison over all 45 cases.

When the values of $L_1$ are considered, the relative importance of these surface parameters for predicting $k_s$ is: $E_x$, $I_x$, $\vert I_z\vert $, $E_z$, $P_o$, $k_{rms}/R_a$, $S_k$, and of least importance, $K_u$. 
The $L_1$-norm error is small when all parameters are included (7.4\%). Excluding any single one of these parameters increases the $L_1$-norm error up to around 9\%.  
On the other hand, the exclusion of $K_u$ from the input parameters does not worsen predictions of $k_s$ significantly. Instead, this observation appears to be a consequence of  correlation between $K_u$ and other surface parameters like $k_{rms}/R_a$ (see figure \ref{fig:pair}). When such correlations exist and one correlating parameter is excluded, the DNN process redistributes the weightings given to other correlated parameters, with little loss in predictive accuracy.

\begin{table}
  \begin{center}{\small 
    \def~{\hphantom{10}}
    \begin{tabular}{L c c c c c c c c c c c c c}
Excluded feature(s) &  
\rotatebox[origin=c]{-45}{None} & 
\rotatebox[origin=c]{-45}{$E_x$} & 
\rotatebox[origin=c]{-45}{$E_z$} &  
\rotatebox[origin=c]{-45}{$E_x,\,E_z$} & 
\rotatebox[origin=c]{-45}{$k_{rms}$} &  
\rotatebox[origin=c]{-45}{$K_u$} &  
\rotatebox[origin=c]{-45}{$k_{rms},\,K_u$} &  
\rotatebox[origin=c]{-45}{$S_k$}& 
\rotatebox[origin=c]{-45}{$P_o$} &  
\rotatebox[origin=c]{-45}{$S_k,\,P_o$} &  
\rotatebox[origin=c]{-45}{$I_x$}&  
\rotatebox[origin=c]{-45}{$I_z$} &  
\rotatebox[origin=c]{-45}{$I_x,\,I_z$} \\
C04	& 2	&-2	& 3	&-1	&-1	&-2	&15	&-1	& 3	&13	&-12	&0	&-3\\
C05	& 5	&-8	&11	& 6	& 3	&\q{-22}	&-4	& 0	& 8	&-4	&-6	&-2	&-11\\
C06	& 0	&10	&-1	& 1	& 0	&10	& 2	& 5	& 1	& 5	&18	& 6	& 8\\
C07	& 1	& 3	&-1	& 2	&10	&\q{-23}	& 0	& 1	&-6	&-1	&13	& 1	& 9\\
C08	&-15	&-14	&-1	&-4	&\q{-19}	&\q{-24}	&\q{-19}	&-2	&\q{-23}	&\q{-36}	&-4	&-7	&-9\\
C09	&18	& 4	& 6	& 3	& 0	& 3	& 6	& 1	&-2	& 6	& 5	&11	& 8\\
C10	&0	& 1	&-16	& 1	&-14	&0	&-1	&-12	& 2	&-13	&15	&0	&11\\
C11	&-12	&-3	&-3	&-23	&-2	&-2	&-5	&-12	&-1	&-29	& 1	&-2	&-2\\
C12	&0	&-4	&-4	& 0	&-18	&-3	&-4	&-1	&-7	&-2	&-3	& 0	&-2\\
C14	& 0	& 4	& 5	& 5	& 1	& 5	&\q{26}	& 3	& 8	&-6	& 3	& 6	& 0\\
C15	&16	& 5	& 0	& 0	& 2	& 9	& 0	& 0	&-11	&-2	& 4	&-5	& 4\\
C16	& 1	&-2	&-1	&\q{24}	&-2	&-2	&-3	& 3	&-2	& 6	& 6	&-1	&14\\
C17	&-4	& 8	&17	&17	& 1	& 4	& 8	&15	&13	&-4	& 3	& 5	& 3\\
C18	&-1	&-6	&-10	&-11	&-2	&-3	&-11	&-3	&\q{-21}	&-17	&-10	&\q{-25}	&-16\\
C19	&-10	&-15	&-11	&-12	&-3	& 4	& 5	& 6	&-4	&-11	&-1	&-2	&-11\\
C20	& 1	& 3	& 3	& 4	& 3	& 3	& 2	& 4	& 3	&0	&\q{23}	&\q{25}	&13\\
C21	& 9	& 2	& 1	& 3	& 1	& 1	& 2	&-1	&0	& 0	& 0	& 8	&14\\
C22	&-3	&-3	&-8	&-9	&-2	&-6	&-8	&-3	&-8	&-9	&-9	&-20	&-12\\
C23	&0	&-2	&-1	&0	&0	&-5	&-17	&-1	& 0	&-1	& 2	&-3	& 2\\
C24	&0	&\q{-21}	&-1	&-1	&-1	& 1	&0	&0	&0	& 4	&0	&-4	&-7\\
C26	&-6	&-17	&-12	&-9	&-8	&-5	&\q{-19}	&-15	&-13	&-5	&-13	&-14	&-10\\
C28	&18	&19	&21	&\q{26}	&17	&18	&-3	&16	&16	&32	&\q{21}	&14	&\q{20}\\
C29	&-9	&-19	&-8	&-22	&-6	&-5	&-13	&\q{-25}	&-11	&-22	&-18	&-17	&-19\\
C30	&-4	& 6	&11	&\q{25}	&-10	&0	& 6	&\q{24}	&0	&-8	& 2	& 6	& 5\\
C31	&\q{22}	&\q{20}	& 8	&19	&\q{24}	& 0	&-2	&18	&-1	&-14	& 9	&-1	& 9\\
C37	&-2	&-8	&-7	&-3	&10	&-4	&-5	&-1	&-5	&-1	&-9	&-8	&-12\\
C40	&-3	&-6	&\q{-27}	&-21	&-6	&-5	&-7	&0	&-1	& 2	&-10	&-8	&-18\\
C43	& 3	&-4	&-4	& 6	&16	& 1	& 2	& 0	& 7	&23	&-15	&-1	&-12\\
C44	&-6	&15	& 1	&17	&13	& 1	& 4	&20	&-6	&-12	&-2	&-16	&\q{-21}\\
C45	& 1	& 2	& 1	&-4	&-6	& 5	&-1	&-11	& 1	& 1	& 5	& 2	& 9\\
E01	&12	& 4	& 4	&-9	& 2	&-3	&-11	& 5	&11	&-10	& 1	&-3	&-3\\
E02	&-13	& 6	&-6	&-7	&-2	&12	& 1	&-2	&10	&-9	&13	& 7	&-2\\
E03	&15	&-6	&0	&-5	& 4	&-6	&-4	& 3	& 7	&-32	& 2	& 1	& 2\\
E04	& 0	&-15	&-9	&-9	&-2	&-6	&-6	&-3	&-5	& 2	&-2	& 4	&0\\
E05	& 5	&17	& 5	&17	& 4	& 9	& 9	& 7	& 5	&28	& 8	& 5	&13\\
E06	&-5	&-3	&-6	&-3	&-10	&-9	&-10	&-6	&-7	&-9	&-10	&-10	&-5\\
E07	&\q{-21}	&\q{-21}	&\q{-24}	&-18	&-16	&-21	&-18	&-17	&\q{-23}	&\q{-41}	&\q{-25}	&\q{-25}	&\q{-24}\\
E08	&\q{22}	&\q{22}	&\q{25}	&22	&\q{19}	&18	&\q{25}	&\q{24}	& 7	&24	&\q{21}	&22	&\q{24}\\
E09	& 5	&-3	&15	&\q{27}	&-1	&\q{22}	&\q{26}	&\q{21}	&-2	&-21	&-3	& 2	& 2\\
E10	&-18	&-19	&-5	&-8	&\q{-25}	&-4	&-5	& 1	&-14	&\q{38}	&-14	& 8	&-2\\
E11	&-1	&-15	&\q{-23}	&-19	&-7	&16	&12	&-2	& 9	&29	& 0	&-5	&0\\
E12	&-9	&-3	& 6	& 0	&-10	& 2	&-2	&-15	&-10	&28	&-15	&-22	&-4\\
E13	&11	& 8	&17	& 6	&17	& 2	& 8	& 7	&\q{21}	&-15	&14	&\q{25}	&15\\
E14	&\q{22}	& 6	& 1	&0	& 6	& 4	& 2	& 1	&\q{25}	&\q{33}	& 9	& 5	&-5\\
E15	&0	&18	&18	&-4	&11	& 9	&15	&11	&19	&32	&19	&23	&16\\\hline
$L_1$	&7.4	&8.9	&8.2	&9.7	&7.6	&7.1	&7.9	&7.3	&8.0	&14.2	&8.8	&8.6	&9.1\\
$L_\infty$	&22	&22	&27	&27	&25	&24	&26	&25	&25	&41	&25	&25	&24\\
    \end{tabular}
     \caption{Errors in $k_s$ prediction by excluding one or two features. The base prediction includes all primary variables. The four largest errors (in magnitude) for each column are colored in red. The errors are percentages. }
     \label{tab:err_sens}}
  \end{center}
\end{table}

To reduce the correlation between the excluded parameters and the remaining ones, one may exclude groups of parameters that are thought to characterize the same type of surface feature.  For this reason, a sensitivity analysis was carried out on the effect of groups of variables on prediction of $k_s$. The characteristics of surface slope, element inclination angle, porosity, and intensity of height fluctuations, are contained in pairs of ($E_x$, $E_z$), ($I_x$, $I_z$), ($P_o$, $S_k$) and ($k_{rms}$, $K_u$),  respectively. Parameters within each pair have been shown to be  correlated to some degree in figure~\ref{fig:pair}. Table \ref{tab:err_sens}   shows  how the accuracy of $k_s$ prediction is affected, if any one of these pairs is excluded. According to the table, the prediction of $k_s$ is sensitive to all four pairs, but with greater sensitivities to the surface porosity (described by $P_o$, $S_k$) and the surface slope (described by $E_x$ and $E_z$). 
As expected, the elimination of both parameters of a pair worsens the prediction more than removing either single parameter (from around 7-9\% errors to up to 14\%).

According  to the sensitivity analysis, all parameters considered are of some importance in the prediction of $k_s$. The effective $x$-slope $E_x$ and roughness height skewness $S_k$ have been suggested as especially significant in earlier studies \citep{NapoliNAD08a,FlackS10,YuanP14a}. The inclination angle in the streamwise direction $I_x$ makes a significant contribution to the $k_s$ prediction because, physically,  $I_x$ characterizes the average aerodynamic shape of the roughness elements. Surfaces with $I_x>0$ are aerodynamically bluff bodies when compared with surfaces of the same size but with $I_x=0$, and surfaces with $I_x<0$ tend to be more streamlined and hence produce less drag. 

An important  finding from this study is that the effective $z$-slope $E_z$ is of similar importance to accurate $k_s$ prediction as $S_k$ or $E_x$.
The exclusion of $E_z$ adversely affects the prediction for a large number of rough surfaces. Physically, $E_z$ describes whether the surface is close to a two-dimensional (2D) roughness with $E_z=0$ (such as a transverse bar roughness) or a three-dimensional (3D) roughness with finite $E_z$. It is known that a k-type 2D roughness produces a higher drag than a 3D roughness with the same height due to the larger spanwise lengthscale that the 2D roughness imparts to the flow~\citep{VolinoSF09}.

\subsection{Comparison between ML algorithms and  polynomial models}  
Explicit algebraic data representations, such as polynomial functions, can also be determined for the data sets of this study, using fitting or minimization procedures. In such methods, a set of basis functions is proposed for a model, the unknown coefficients of which are then optimized according to specified constraints. They are a generalization of the models of equation (\ref{eq:empmdls}), which were based on experimental observations of the dependence of $k_s$ on a small number of surface parameters.  A 30-degree-freedom polynomial basis was proposed as a `white-box' model for $k_s$,  analogous to a low-order Taylor series expansion for $k_s$:
\begin{equation}
\begin{aligned}
     k_s/R_a  =\  & \alpha_0 + 
         \alpha_1 (k_{rms}/R_a)^{\alpha_2}   + \alpha_3 I_x + \alpha_4 \vert I_x\vert^{\alpha_5}  + \alpha_6 \vert I_z\vert + \alpha_7 \vert I_z\vert^{\alpha_8}+ \\
         & \alpha_9 P_o^{\alpha_{10}} + \alpha_{11}E_x^{\alpha_{12}} + \alpha_{13}E_z^{\alpha_{14}} + \alpha_{15}S_k + \alpha_{16}\vert S_k\vert^{\alpha_{17}} +\\ & \alpha_{18}(K_u-3) + \alpha_{19}\vert K_u-3\vert ^{\alpha_{20}} 
         +\alpha_{21}(k_{rms}/R_a)^{\alpha_{22}}P_o^{\alpha_{23}}
         +
         \\&\alpha_{24}(k_{rms}/R_a)^{\alpha_{25}}E_z^{\alpha_{26}}
         +\alpha_{27}P_o^{\alpha_{28}}E_z^{\alpha_{29}},
\end{aligned}\label{eq:wht}
\end{equation}
where $a_i$ ($i=0,1,\cdots, 29$) are the model coefficients.  To keep this model as simple as possible and to bring the effects of all contributing factors into account, we used terms as $\alpha_i \theta ^{\alpha_j}$ for a test variable $\theta$ that take only positive values (e.g. $k_{rms}$), and terms as $\alpha_i \theta + \alpha_j|\theta| ^{\alpha_k}$ for those variables that take both positive and negative values (e.g. $S_k$). For the latter,  the power of $\theta$ in the first term is fixed (at one) instead of fitted, to eliminate the possibility of an imaginary number. Combinations of six parameters ($E_x$, $E_z$, $P_o$, $S_k$, $k_{rms}/Ra$ and $K_u$), taken in pairs,  were also included. Since, for the present collection of surfaces, strong correlations were observed between individual variables within the  three pairs of $(E_x,\,E_z)$, $(P_o,\,S_k)$ and $(k_{rms}/Ra,\,K_u)$, shown in figure~\ref{fig:pair}, only one variable from each pair was used for the combination terms in equation~(\ref{eq:wht}). Using the other variable from any of these pairs instead would not lead to a significant change in the prediction using equation~(\ref{eq:wht}).  

The high-dimensional space of $a_i$ is poorly suited to curve-fitting and minimization procedures which use stochastic gradient descent algorithms. However, it is well suited to robust minimization methods like the differential evolution algorithm \citep{Storn97}, with which global minima can often be found efficiently in spaces of high dimension. In this case, it is used to determine the values of the coefficients $a_i$ which minimize the $L_1$ norm. 

\begin{figure}\centering
    \includegraphics[width=.8\textwidth]{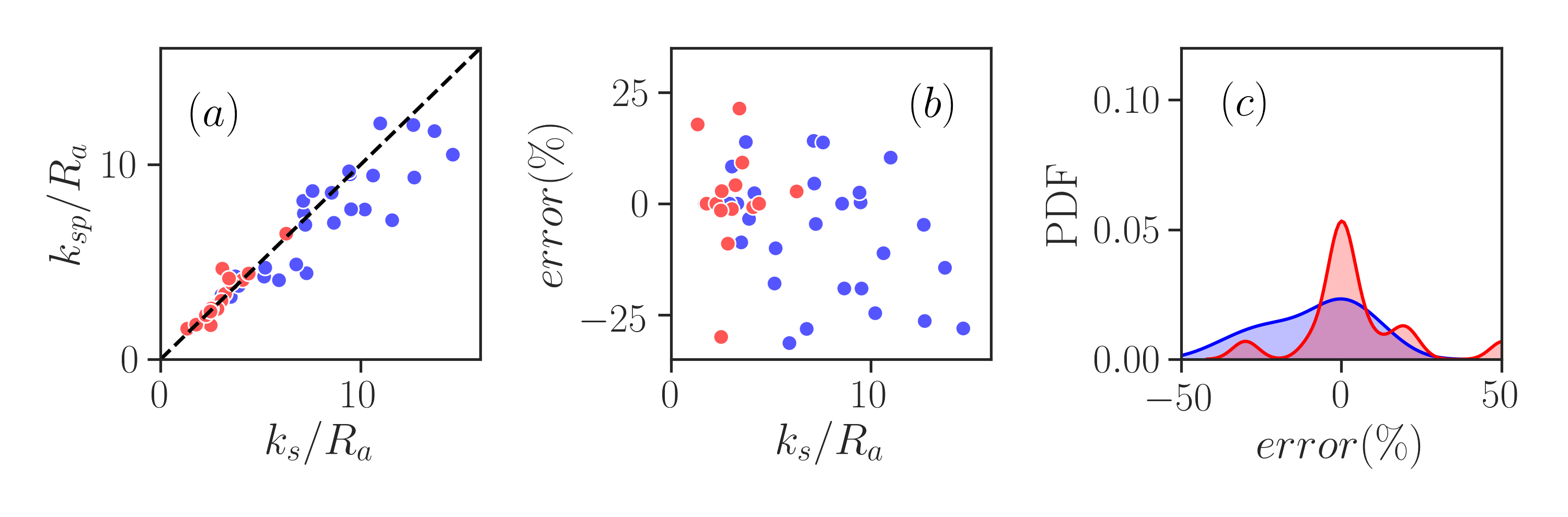}
    \caption{   (a) Scatter plot of true $k_s$ and predicted $k_s$ (denoted as $k_{sp}$),  (b)  scatter plot of  true $k_s$ and    relative error and (c) pdf of relative error distribution for prediction using  polynomial function defined in equation~(\ref{eq:wht}), with DNS data (\textit{blue}) and experimental data (\textit{red}).} 
    \label{fig:err_DE}
\end{figure}

In figure \ref{fig:err_DE}, the prediction quality of this white-box model with  optimized coefficient values is shown. This method yields an average prediction error of 12\% and a maximum one of 51\% when using all 45 fully-rough data sets (to give the best possible prediction accuracy) for the model training.

The optimized values of $a_i$'s are
\begin{center}
\begin{tabular}{l l l l l l}
$\alpha_0$~~=~  5.312, & 
$\alpha_1$~~=  -1.172,& 
$\alpha_2$~~=~  4.264, & 
$\alpha_3$~~=~  0.050,& 
$\alpha_4$~~=  -1.283,& 
$\alpha_5$~~=~  8.393,\\
$\alpha_6$~~=  -0.347, & 
$\alpha_7$~~=  -5.771,& 
$\alpha_{8}$~~=~ 1.785,& 
$\alpha_{9}$~~=~ 7.919, & 
$\alpha_{10}$~=~ 4.058, & 
$\alpha_{11}$~=  -0.979, \\
$\alpha_{12}$~=~  3.414, &
$\alpha_{13}$~=~  6.380,& 
$\alpha_{14}$~=~  1.354,&
$\alpha_{15}$~=~  1.023, & 
$\alpha_{16}$~=~  2.969,& 
$\alpha_{17}$~=~  1.273,\\
$\alpha_{18}$~=  -0.946, &
$\alpha_{19}$~=  -0.762, & 
$\alpha_{20}$~=~  0.056,& 
$\alpha_{21}$~=~  1.647,  & 
$\alpha_{22}$~=  -8.176,  &
$\alpha_{23}$~=~  3.523, \\
$\alpha_{24}$~=  -9.472,  &
$\alpha_{25}$~=  -5.656,  & 
$\alpha_{26}$~=~  0.580, &
$\alpha_{27}$~=  -5.425, & 
$\alpha_{28}$~=~  0.283,& 
$\alpha_{29}$~=~  7.177.  
\end{tabular}
\end{center}

The predictive accuracy of this optimized explicit model equation is considerably lower than that of the DNN and GPR methods. One reason for this reduced accuracy is that low-order functions of geometrical parameters do not faithfully represent the dependence of $k_s$ on surface parameters because each coefficient in the model is required to take the same value over the entire surface-parameter space.  In ML approaches, such restrictions need not apply as they are not constrained to low-order polynomial functions but instead adopt a methodical search for the best representation of $k_s$ as a function of the surface parameters. This search is carried out through `feature selection' in the first layers of DNN and the properties of the basis functions adopted in GPR, each of which are designed to yield the same mean and standard deviation of $k_s/R_a$ as in the original dataset  \citep{Rasmussen06}. 

\section{Concluding remarks}\label{sec:conclusion}

The construction of a predictive model from a large ensemble of dataset for the
equivalent sandgrain height $k_s$ of a surface of arbitrary roughness, as a
function of many different measures of surface topography, is a labeled
regression problem that is well-suited to machine learning techniques. In this
paper, data from 45 different rough surfaces (in fully rough flows) were used to devise DNN and GPR predictions for $k_s$ as functions of 8
different surface-roughness parameters.

Both models were able to predict $k_s$ for the 45 surfaces with an average error below 10\%, with the largest error for any one
surface less than 30\%.  These predictions were significantly better than those
of existing formulas, and of a 30 degree-of-freedom polynomial model fitted to
the same data, where the greatest error for any surface was about 50\%. 

Sensitivity analyses revealed that inclusion of nearly all the surface roughness descriptive parameters was necessary to minimize  the
average prediction error, but that exclusion of either measures of porosity or measures of the surface slope increased the maximum prediction error more significantly than omitting other parameters.

Machine learning techniques are well suited to this modeling problem because:
\textit{i}) it is complex insofar as different kinds of surface roughness yield
different flow phenomena which are modeled most accurately in different ways,
making the prospect of a general physical model very remote; and \textit{ii})
the dependent surface-roughness variables upon which $k_s$ is modeled are a
large non-orthogonal set for which robust multivariable regression techniques
are required.  As machine learning methods, they take no account of physical
modeling concepts or observed phenomena within roughness sublayers, such as
recirculation regions, enhanced turbulence production in the wake of roughness
elements, assumed scalings for drag \textit{etc.}, each of which is applicable
to flows over some rough surfaces but not others.  Nor are they hindered by the
lack of orthogonality of the surface roughness parameters as the dependent
variables of $k_s$.  The techniques used can be configured readily to mimic
models with very many degrees of freedom and, when compared to polynomial
models, their feature selection properties provide the equivalent of different
values for polynomial coefficients in different regions of the surface-parameter
space.	In this application, both approaches of DNN and GPR yielded models with very similar
predictive accuracy, even though the techniques themselves were very different.
We therefore conclude that they yield high-fidelity predictions of the
equivalent sand-grain roughness height for turbulent flows over a wide range of
rough surfaces, as a significant improvement over other methods. Improved prediction might be achieved by enlarging the database to include rough-wall flows with surface parameters which correspond to the relatively low prediction confidence in the GPR method, and by including
additional roughness parameters as inputs which might describe sparseness and two-dimensionality, such as the solidity, correlation lengthscales and other two-point surface statistics. 

In addition to the $k_s$ prediction described here, the DNS database and the ML techniques in general can also be used to uncover  relations between roughness geometry and physics-related quantities, such as the flow pattern around roughness protuberances, flow separation locations, characteristics of  the shear layers associated with the separation bubbles, the wake sheltering volume, \textit{etc.}  Specifically, a ML network trained to correlate these flow characteristics (as outputs) to the roughness geometry (as inputs) may be an efficient tool for determining the sets of  roughness geometrical features which are important for characterizing these effects. Knowledge of such a set of significant roughness parameters may also guide the construction of rough-surface databases that yield more efficient and more widely applicable predictions of $k_s$ or other quantities. 

\section{Declaration of interests}
The authors report no conflict of interest.

\section{Acknowledgements}

The authors gratefully acknowledge the financial support of the Office of Naval Research (Award No. N00014-17-1-2102). Computational support was provided by Michigan State University's Institute for Cyber-Enabled Research. The authors also gratefully thank Professor Karen A. Flack of the US Naval Academy for providing the experimental data sets.   

\section{Supplementary materials}\label{sec:suply}
The rough-wall flow database (including  $k_s$, surface height map and surface parameters)  and the trained DNN and GPR networks, called Prediction of the Roughness Equivalent Sandgrain Height (PRESH), can be accessed online in the first author's GitHub repository at \href{https://github.com/MostafaAghaei/Prediction-of-the-roughness-equivalent-sandgrain-height}{https://github.com/MostafaAghaei/Prediction-of-the-roughness-equivalent-sandgrain-height}.
With this package of data and programs, interested researchers can: i) use the ML networks described in this paper to make  predictions of $k_s$ for surfaces of their own  roughness topography; ii) download the code and train new DNN and GPR networks to predict $k_s$ for a different set of surfaces of arbitrary topography; and iii) use the database of 45 rough-wall flows for other  applications. 
It is recommended to use the ML configurations described in this paper for surfaces with  parameters inside the ranges specified in figure \ref{fig:pair}. Extrapolations (using inputs which are beyond the specified range) will lead to additional uncertainty. 

The PRESH and the database will be actively updated by the authors to improve the prediction  accuracy and universality. We welcome  interested researchers to share their datasets with us.

\bibliographystyle{jfm}
\bibliography{Biblio}

 \end{document}